\documentclass[a4paper,fleqn]{cas-sc}

\usepackage[authoryear]{natbib}
\usepackage{enumitem}
\usepackage[capitalize]{cleveref}
\usepackage{svg}
\usepackage{float}
\usepackage{booktabs} 
\usepackage{multirow} 
\usepackage{makecell} 
\usepackage{tabularx} 

\hypersetup{
    colorlinks=true,
    citecolor=blue,
    linkcolor=blue,
    urlcolor=blue
}

\AtBeginDocument{\hypersetup{citecolor=blue}}   

\begin{document}
\setlength{\abovedisplayskip}{6pt}
\setlength{\belowdisplayskip}{6pt}
\setlength{\abovedisplayshortskip}{4pt}
\setlength{\belowdisplayshortskip}{4pt}

\let\WriteBookmarks\relax
\def\floatpagepagefraction{1}
\def\textpagefraction{.001}

\shorttitle{CV Data-based Arterial Network Traffic State Prediction: AASTGCN}
\shortauthors{Han et~al.}

\title [mode = title]{Arterial Network Traffic State Prediction with Connected Vehicle Data: An Abnormality-Aware Spatiotemporal Network}

\author[1]{Lei Han}[orcid=0000-0002-2976-0762]
\ead{lei.han@ucf.edu}
\cormark[1]
\author[1]{Mohamed Abdel-Aty}[orcid=0000-0002-4838-1573]
\ead{M.aty@ucf.edu}
\author[1]{Yang-Jun Joo}[orcid=0000-0003-2319-0874]
\ead{yangjun.joo@ucf.edu}
\affiliation[1]{%
  organization={Department of Civil, Environmental and Construction Engineering, University of Central Florida},
  city={Orlando},
  postcode={32816},
  state={FL},
  country={United States}
}
\cortext[1]{Corresponding author}

\begin{highlights}
    \item Leverage high-resolution CV data to estimate arterial network traffic state measures (e.g., travel delay and queue).
    \item Propose an Abnormality-Aware Spatiotemporal Graph Convolution Network (AASTGCN) to predict arterial-network delay and queue length. 
    \item AASTGCN adopts two spatiotemporal branches with gated fusion to capture both real-time traffic dynamics and long-term traffic periodic patterns.
    \item A dual-expert architecture is incorporated to model normal and abnormal traffic separately, disentangling their distinct patterns and improving prediction accuracy under both conditions.
    \item Experiments on a city-scale arterial network with 1,050 links validate the SOTA performance of AASTGCN.
\end{highlights}

\begin{abstract}
Emerging connected-vehicle (CV) data shows great potential in urban traffic monitoring and forecasting. However, prior CV-based studies on arterial traffic measures prediction (i.e., delays and queues) are limited to simulated high-penetration scenarios or small networks, which are challenging to apply in real-world city-scale arterial networks. To address such gaps, we develop a CV data-based arterial traffic prediction framework with two components: (1) a two-stage traffic state extraction method that estimates vehicle-level traffic measures from CV trajectories and then aggregates them into network-level traffic state measures; (2) an Abnormality-aware spatiotemporal graph convolution network (AASTGCN) that adopts a dual-expert architecture to separately model normal and abnormal traffic, and jointly captures short-term traffic dynamics and long-term periodicity via spatiotemporal GCN with a gated-fusion mechanism. Real-world CV data are used to test our method in a large arterial network with 1,050 links (386.4 miles). Experimental results show that: 1) the proposed traffic estimation method is effective for large arterial networks to provide real-time traffic measures (e.g., link-level average travel delay and queue length), which are critical for urban traffic operation and evaluation. 2) Abnormal traffic prediction is typically challenging for existing methods. By modeling abnormal cases separately from normal traffic in two dedicated experts, AASTGCN outperforms existing models for both normal and abnormal traffic conditions. 3) The gate-fusion mechanism adaptively balances real-time and historical information: it leverages more historical-periodic information in normal traffic and shifts a higher weight to real-time traffic dynamics for abnormal traffic deviating abruptly from historical patterns. The proposed framework can be implemented in large urban arterial networks for various real-time traffic operation applications.
\end{abstract}


\maketitle

\section{Introduction}
Arterial networks carry a substantial volume of urban traffic and serve as the backbone of urban transportation systems. 
Reliable estimation and prediction of traffic states on these networks are critical for traffic operations and management applications \citep{Zhan2020Link}. 
Traditional traffic state measuring approaches have relied on fixed sensors such as loop detectors \citep{Liu2007Estimation,Wang2012Comparison}, automated vehicle identification sensors \citep{Li2011Incorporating}, microwave vehicle detection systems \citep{Shi2015BigData,Shi2016Evaluation} and license-plate recognition cameras \citep{Zhan2020Link,He2024Vehicle}. 
However, the high installation and maintenance costs typically restrict their deployment to specific road types or small networks. Consequently, most of these studies are conducted on expressways \citep{Shi2015BigData,Shi2016Evaluation,Abdelraouf2022Utilizing,Zhang2025CASAformer} rather than urban arterials.
Although a few studies have examined arterial roads, their study areas are typically limited to a few arterial segments or networks \citep{Zhan2020Link,He2024Vehicle}. 
These data constraints limit the applicability of traditional approaches to real-world, city-scale arterial networks. 

With advances in vehicle communication techniques, high-resolution connected vehicle (CV) data have become a promising data source for urban traffic monitoring  \citep{Wang2024aTraffic}. 
Unlike fixed sensors, CV data are obtained from individual vehicles without spatial restrictions, which can cover large-scale urban road networks and diverse road types. 
Therefore, recent studies have shifted towards utilizing CV data for urban arterial traffic state analysis \citep{Ghandeharioun2022Link,Subramaniyan2023Hybrid,Novak2023DataDriven,islam2024signal,Shafik2025RealTime}. 
For example, travel time and delay can be derived from CV arrival and departure times and are widely used as key indicators of arterial traffic performance.
Substantial efforts have focused on estimating and predicting these measures \citep{Chen2018Exploring,Subramaniyan2023Hybrid,Roy2025MultiTask}. 
Queue length is another critical indicator to measure arterial traffic effectiveness. 
As it is difficult to estimate from stationary sensors, numerous queue-length estimation and prediction methods using CV trajectories have been proposed \citep{Zhao2019Estimation,Comert2021Grey,Tan2021CycleBased}. 
However, most of the existing studies rely on simulated virtual data and assume relatively high penetration rates (e.g., 10-50\%) \citep{Tavafoghi2021Queue}. 
Only a few studies validated their methods with real-world cases \citep{Tan2021CycleBased,Shi2023Queue}, yet their samples are generally limited to specific arterial segments and networks. 
To the best of our knowledge, large-scale arterial traffic state estimation and prediction of delay and queue length using real-world CV data remains an open research gap. 

Motivated by the above gaps, this study aims to estimate and predict segment-level delay and queue length over a city-scale arterial network with high-resolution real-world CV data. 
The key to accurate prediction of these measures relies on capturing complex temporal patterns and spatial dependencies in arterial traffic. 
To this end, deep learning (DL) models, particularly spatiotemporal graph networks, have been widely adopted due to their superior ability to model such spatiotemporal dependencies \citep{Jiang2023PDFormer,Novak2023DataDriven,Li2024STABC,Zhang2025CASAformer}. 
However, prior methods still face two limitations. 
First, most studies primarily extract short-term traffic dynamics to forecast future traffic without accounting for the long-term temporal periodicity of arterial traffic (e.g., recurring patterns on the same day of the week). 
Failing to incorporate such periodic information may hinder model generalization and limit prediction accuracy \citep{Abdelraouf2022Utilizing}. 
Second, abnormal events (e.g., crashes and non-recurrent congestion) may substantially disrupt typical traffic patterns, yet most DL frameworks fail to differentiate these abnormal conditions, modeling them mixed with normal traffic \citep{Guo2010Comparison,Xu2021Flexible}. 
Given that abnormal traffic states are relatively rare, such mixed training will bias the model toward dominant normal traffic, resulting in poor robustness on sparse yet critical abnormal events and ultimately degrade overall prediction performance. 

To address these challenges, we propose an Abnormality-Aware Spatiotemporal Graph Convolution Network (AASTGCN) for predicting arterial network delay and queue length. 
The main contributions of this paper are summarized as follows:

\begin{itemize}[nosep]
    \item Leveraging high-resolution CV data, a two-stage traffic state extraction framework is developed that first estimates vehicle-level traffic measures and then aggregates them to obtain link-level traffic state measures across arterial networks.
    \item Within AASTGCN, two spatiotemporal branches are introduced and fused via a gated mechanism to simultaneously capture real-time traffic dynamics and long-term traffic periodic patterns from historical observations.
    \item Abnormal traffic states are explicitly detected based on historical traffic distributions. Dual-expert architecture is employed for normal and abnormal traffic to disentangle their distinct patterns and improve prediction accuracy under both conditions.
    \item Experiments on a city-scale arterial network with over 1,000 links (386.4 miles) validate the state-of-the-art (SOTA) performance of the proposed AASTGCN.
\end{itemize}

Following this section, \cref{sec:literature} reviews relevant literature.
\cref{sec:extraction} details the CV-data based traffic state extraction, and \cref{sec:aastgcn} introduces the framework of AASTGCN.
\cref{sec:experiments} provides experiment design and results. Finally, \cref{sec:conclusion} presents the conclusion.

\section{Literature Review}
\label{sec:literature}
\subsection{CV Data-based Traffic State Estimation}
Owing to their wide spatial coverage and vehicle-level granularity, CV data have emerged a primary data source for urban arterial traffic state estimation. 
A series of traffic measures (e.g., speed, travel time, delay, and queue length) can be derived from CV trajectories to reflect traffic states on arterial networks \citep{ArgoteCabanero2015Connected,Wang2023Trajectory}. 
Among these measures, travel time and delay are more readily derived from CV data, and most of the existing studies have focused on their estimation. 
For example, \citet{Jenelius2013Travel} proposed a statistical model for urban road network travel time estimation using low-frequency probe vehicle GPS data; \citet{ArgoteCabanero2015Connected} conducted simulation-based experiments for a major arterial in Berkeley, CA, showing that a CV penetration rate as low as 1\% could provide representative estimates of travel time. 
Using the NGSIM dataset, \citet{Iqbal2019Effect} simulated different CV penetration scenarios and claimed that 3–4\% penetration on urban streets can yield reliable travel time estimation.
\citet{Correa2024Urban} propose a data-driven matching-algorithm to achieve fast link travel time estimation from high-sampling-rate CV data at the city scale.

Queue length is another key performance measure for arterial traffic operations, and a big part of recent work has investigated CV data-based queue length estimation. 
For instance, shockwave-based methods have been proposed to identify critical queue-related events (e.g., joining and leaving the queue) from CV trajectories and then estimate queue length \citep{Ban2011Real,Cheng2012Exploratory,Li2017RealTime,Yin2018Kalman}. 
Several studies have developed stochastic frameworks to estimate queue length using the observed locations of CVs at each cycle \citep{Comert2013Simple,Comert2016Queue,Comert2024Simple,Zhao2019Estimation}. 
Other stochastic estimation methods infer queue length by modeling arrival and departure processes at the intersection approach \citep{Hao2014Cycle,Tiaprasert2015Queue,Tan2021CycleBased}. 
Despite these advances, most queue-length estimation methods assume relatively high CV penetration rates (10\%–50\%), which is unrealistic in current real-world deployments.
Additionally, most algorithms require detailed knowledge, such as signal timing and CV penetration rate, making them difficult to deploy at scale. 
As a result, these methods remain primarily confined to simulation-based settings or limited arterial segments and networks.  

Overall, toward real-world city-scale arterial networks, how to effectively estimate traffic state measures (e.g., delay and queue length) with emerging CV data and further enable real-time prediction remains largely unexplored and warrants further investigation. 

\subsection{Traffic Prediction Deep Learning Methods }
Given that traffic prediction is inherently a spatial-temporal forecasting task, DL models have been widely adopted for their powerful capabilities to extract intricate spatiotemporal correlations and achieved superior prediction performances in recent years. 
Initially, recurrent neural networks (RNNs) and its variants (e.g., long and short-term memory (LSTM) and gated recurrent unit (GRU) networks) are employed to model temporal correlations of traffic features \citep{Tian2015Predicting,Zhang2017Combining}. 
Convolutional neural networks (CNNs) are introduced to capture spatial dependences \citep{Zhang2018Predicting,Zheng2021Hybrid}. 
However, while CNNs are effective for grid-structured inputs, they can hardly handle the inherent graph structure of road networks. 
This non-Euclidean characteristic of traffic networks imposes significant limitations on the efficacy of CNNs in handling the spatial dependencies and topological constraints in traffic data \citep{Fan2025PDG2Seq}.

Conversely, graph neural networks (GNNs) have demonstrated strong abilities in modeling graph-structured traffic data by explicitly encoding network topology and neighborhood-based information with flexible receptive fields \citep{Khaled2022TFGAN,Liu2022GraphSAGE}. 
Researchers combined GNNs with temporal modules (e.g., RNNs) as spatiotemporal graph neural networks (STGNNs) to jointly capture spatial and temporal dependencies in traffic flow and achieve improved prediction performances \citep{Li2018Diffusion,Yu2018Spatio}. 
Subsequently, a variety of STGCN-based models—such as Graph WaveNet \citep{Wu2019Graph}, ASTGCN \citep{Guo2019Attention}, T-GCN \citep{Zhao2020TGCN}, GMAN \citep{Zheng2020GMAN}, STGIN \citep{Zou2023Novel}, STABC \citep{Li2024STABC} and PDG2Seq \citep{Fan2025PDG2Seq}—have been developed to explore spatiotemporal relationships from different perspectives. 
Recently, Transformer-based models have gained popularity and demonstrated strong performance, owing to their attention mechanisms that effectively learn the important spatiotemporal traffic data \citep{Liu2023Spatio,Zou2024ShortTerm,Ouyang2025Graph,Zhang2025CASAformer}. 
For instance, \citet{Chen2023Bidirectional} proposed a bidirectional spatial–temporal adaptive transformer that adopted an encoder–decoder architecture with both spatial-adaptive and temporal-adaptive transformers for accurate traffic forecasting. 
STAEformer \citep{Liu2023Spatio} leverages novel spatiotemporal and adaptive embeddings to learn spatiotemporal patterns in traffic data, providing an effective framework for improving the performance of traffic speed and flow prediction. 



Although existing studies have developed advanced DL models for traffic prediction, two key challenges persist: 
(1) most studies primarily extract short-term traffic dynamics to forecast traffic, while underutilizing critical temporal periodicities (e.g., weekly patterns).  
While some studies have included daily and weekly historical traffic, they often treat these periodic inputs as additional features and simply concatenate them into input temporal sequences, ignoring the distinct distributions between current and historical traffic flow \citep{Zou2026MultiGraph}. 
(2) abnormal traffic prediction remains insufficiently addressed. 
Normal traffic is generally easier to predict as it tends to follow recurring and stable patterns. 
In contrast, abnormal traffic affected by incidents (e.g., crashes or non-recurrent congestion) may substantially disrupt typical traffic dynamics to make it harder to forecast—yet accurate prediction under such conditions is more critical for traffic operation management \citep{Salamanis2017Identifying}. 
However, existing studies typically train models without distinguishing abnormal traffic from normal traffic. 
Given that abnormal traffic states are relatively rare, such training may bias models toward dominant normal traffic, thus reducing robustness to abnormal traffic. 
In summary, effectively incorporating temporal periodicities from historical traffic and enhancing the prediction performance of abnormal traffic conditions remain open challenges, which constitute the primary focus of this study.

\section{CV Data-based Arterial Traffic State Extraction}
\label{sec:extraction}
\subsection{CV Data Processing }
\begin{figure}
    \centering
    \includegraphics[width=0.73\textwidth]{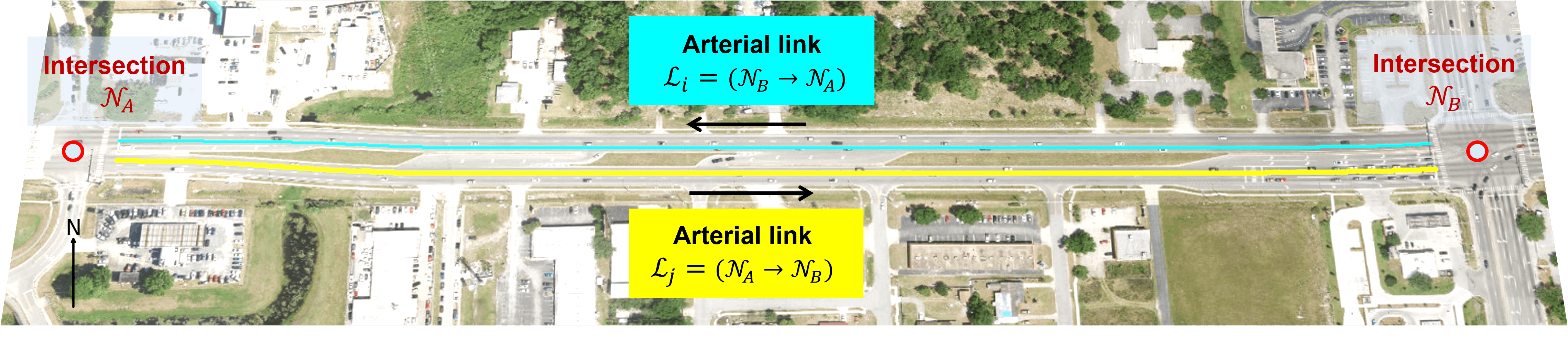}
    \vspace{-1em} 
    \caption{Urban arterial network representation.}
    \label{fig:1}
\end{figure}

The arterial basemap is the foundation for representing the arterial network structure. 
Following common definitions in prior studies \citep{Wang2023Trajectory,Wang2024aTraffic,Zhan2020Link}, arterials are segmented into directional links delineated by intersections and links are adopted as the basic analysis unit.
As a result, the arterial network can be represented as a directed graph
$\mathcal{G} = \{\mathcal{N}, \mathcal{L}\}$,
where $\mathcal{N}$ and $\mathcal{L}$ are the sets of intersections (nodes) and directional links, respectively. 
As shown in \cref{fig:1}, two links $\mathcal{L}_i$ and $\mathcal{L}_j$, between intersections $\mathcal{N}_A$ and $\mathcal{N}_B$ represent two opposite directions. 
This segmentation guarantees the continuity of CV trajectories within each link from the upstream to downstream intersections. 
To achieve an accurate network representation in the basemap, we use OpenStreetMap as raw map data and identify intersections. 
Finally, arterials are manually split into basic road links and refined to ensure that each link spans from the upstream stop line to the downstream stop line. 

Raw connected-vehicle (CV) trajectories are represented as
$Tr_v=\{\mathcal{P}_1,\mathcal{P}_2,...,\mathcal{P}_{n_v}\}$,
where $n_v$ is the number of CV points for vehicle $v$.
Each CV point is defined as
$\mathcal{P}_i=\{\mathrm{lon}_i,\mathrm{lat}_i,t_i,v_i,h_i\}$,
including longitude $\mathrm{lon}_i$, latitude $\mathrm{lat}_i$, timestamp $t_i$,
speed $v_i$ (mph), and heading $h_i$ (degree).

\begin{figure}
    \centering
    \includegraphics[width=0.73\textwidth]{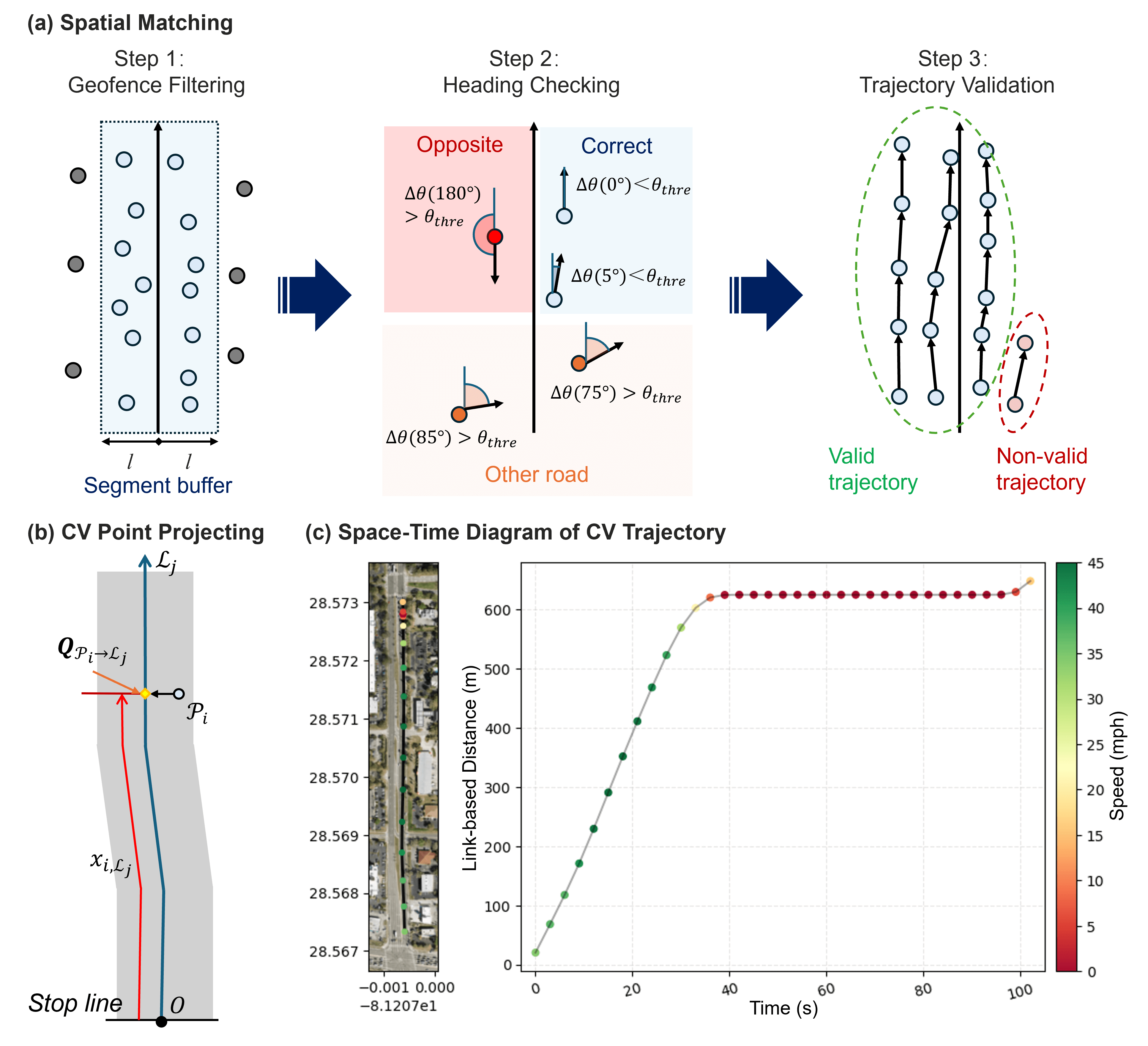}
    \vspace{-1em} 
    \caption{Illustration of CV data processing.}
    \label{fig:2}
\end{figure}

Given the arterial basemap and CV trajectories, CV data processing aims to (i) match CV points to arterial links and (ii) convert the original GPS coordinates to link-based distances. As shown in \cref{fig:2}, two key steps are conducted:

\textbf{Spatial matching.} As shown in \cref{fig:2}(a), each CV trajectory point is first aligned to arterial links through a three-step procedure. 
\textit{Step 1, Geofence filtering:} For each link, a buffer (geofence) with a fixed width $l$ is generated, and CV points within the buffer are selected as candidate groups. 
\textit{Step 2, Heading checking:} To filter out candidate CV points that are not traveling along the target link, the angular deviation $\Delta\theta$ between vehicle heading and the link traveling direction is computed. 
CV points with $|\Delta\theta|<\theta_{\mathrm{thre}}$ are treated as correctly aligned (blue dots), while the remaining points are removed while the remaining points are removed including the opposite CVs (red dots) and CVs on other roads (orange dots).
\textit{Step 3, Trajectory validation:} To filter out vehicles merging from the middle of the road or exiting from the roadside, a journey is considered invalid and removed if it contains fewer than two matched points or its total travel distance is less than half of the link length.
Only journeys satisfying both criteria are retained as the final matched trajectories for each link to obtain a reliable traffic state indicator.

\textbf{CV point projecting.} This step converts the matched CV point from original GPS coordinates (latitude/longitude) to the distance along the link. 
As shown in \cref{fig:2}(b), a CV point $\mathcal{P}_i$ is projected onto link $\mathcal{L}_j$ to obtain the projected point $\mathcal{Q}_{\mathcal{P}_i\rightarrow \mathcal{L}_j}$. 
Then, the distance from the link origin $O$ (upstream stop line) to $\mathcal{Q}_{\mathcal{P}_i\rightarrow \mathcal{L}_j}$ is calculated as the link-based distance of $\mathcal{P}_i$ on $\mathcal{L}_j$, denoted by $x_{i,\mathcal{L}_j}$.

After these two steps, each raw CV point is augmented with two additional attributes: its matched link $\mathcal{L}_j$ and its link-based distance $x_{i,\mathcal{L}_j}$. Thus, $\mathcal{P}_i$ can be represented as $\{\mathrm{lon}_i,\mathrm{lat}_i,t_i,v_i,h_i,\mathcal{L}_j,x_{i,\mathcal{L}_j}\}$.
Therefore, the space-time diagram of a given CV trajectory on each link can be obtained, as shown in \cref{fig:2}(c), serving as the basis for subsequent traffic measure computing.

\subsection{Traffic State Measure Extraction}
With the processed CV data, a two-stage traffic state extraction framework is developed that first estimates single vehicle-level traffic state measures and then aggregates them to obtain link-level measures across the arterial network.

\subsubsection{Single vehicle traffic state measures}
\begin{figure}
    \centering
    \includegraphics[width=0.71\textwidth]{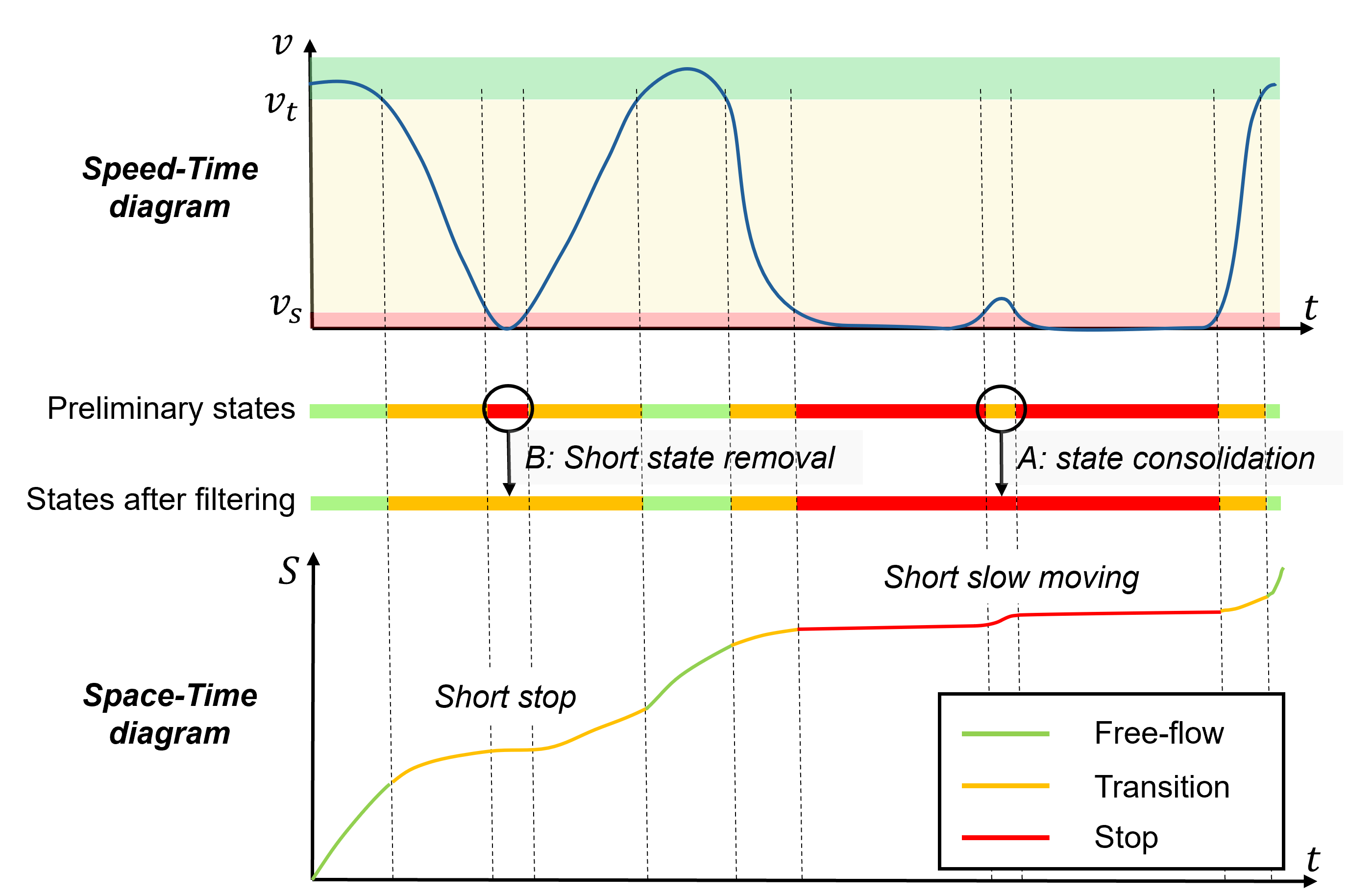}
    \vspace{-1em} 
    \caption{Illustration of trajectory state segmentation.}
    \label{fig:3}
\end{figure}

For each vehicle traversing a link, a trajectory state segmentation is first used to split its entire trajectory into three states: free-flow, transition, and stop, as illustrated in \cref{fig:3}. 
Preliminary trajectory states are first labeled by speed: \emph{Stop} if $v_i \le v_s$ with $v_s=1~\mathrm{m/s}$; \emph{Free-flow} if $v_i>v_t$, where $v_t=0.8\times$ (speed limit) following \citet{Wang2023Trajectory}; and \emph{Transition} otherwise (i.e., $v_s < v_i \le v_t$). 
State filtering is then applied to ensure the segmentation robustness. 
Specifically, \emph{state consolidation} merges two nearby states of the same category if both the time and distance gaps are below predefined thresholds. 
For example, at Point A in \cref{fig:3}, two stop states separated by a short transition state are merged into a longer stop state. 
In this study, the duration threshold is set to $9~\mathrm{s}$ and the distance threshold is $10~\mathrm{m}$. 
In addition, \emph{short-state removal} converts short stop states into transition states, as shown at Point B. 
If the duration of a stop state is less than $9~\mathrm{s}$, it will be converted to a transition state. 
After applying state consolidation and short-state removal, the same filtering procedure is performed for free-flow states. 
The final segmentation results can be obtained as shown at the bottom of \cref{fig:3}. 

\begin{figure}
    \centering
    \includegraphics[width=0.71\textwidth]{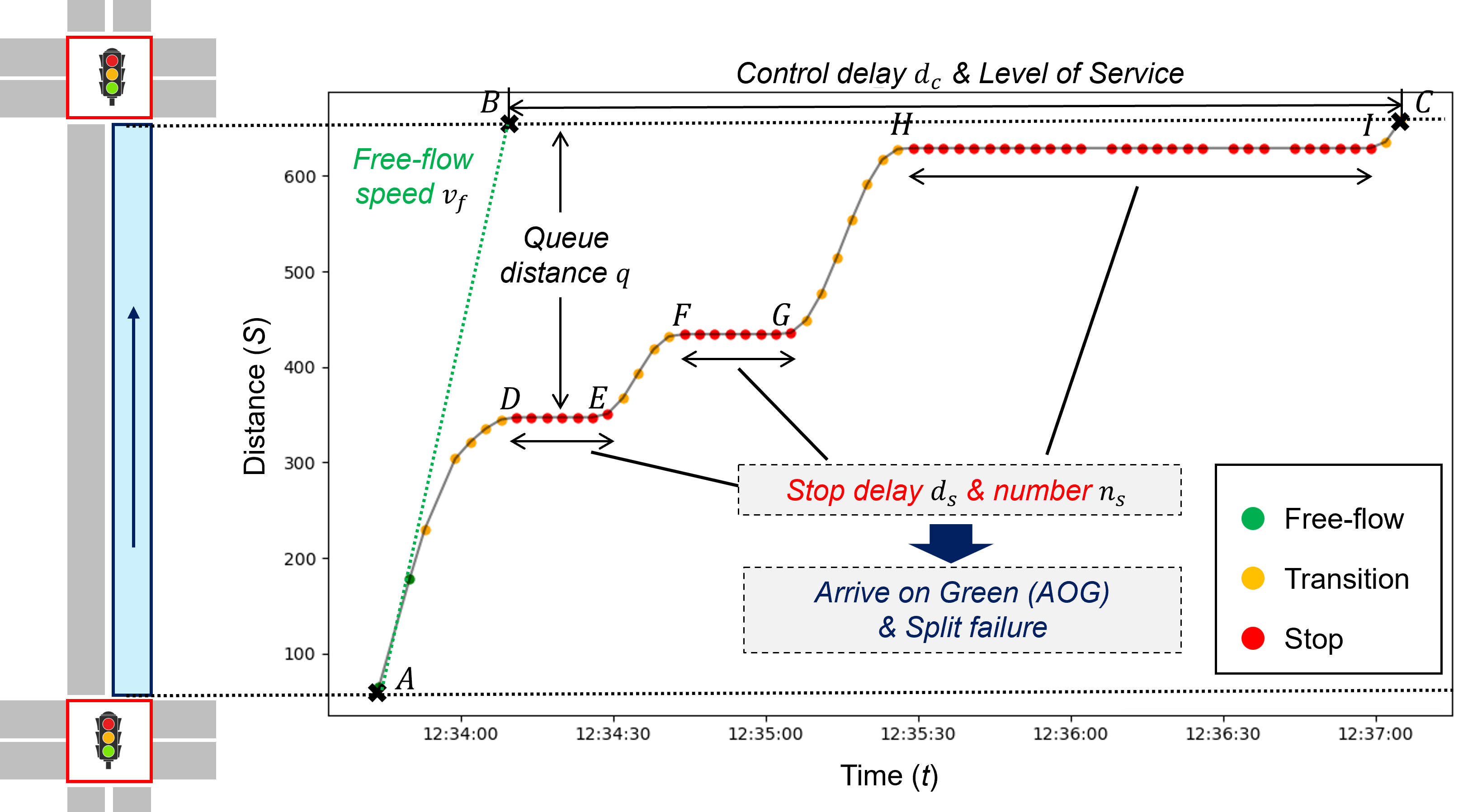}
    \vspace{-1em} 
    \caption{Singel CV traffic state measures computation.}
    \label{fig:4}
\end{figure}

With the segmented trajectories, a series of traffic state measures can be computed from the space--time diagram (see \cref{fig:4}). \Cref{tab:2} summarizes these measures following the procedure in \citet{Wang2023Trajectory}.

\begin{table}
    \centering
    \caption{Summary of single CV traffic state measures.}
    \label{tab:2}
    \footnotesize\rmfamily
    \renewcommand{\arraystretch}{1.2}

    \begin{tabularx}{\linewidth}{p{2.7cm} X p{4.8cm} p{0.8cm}}
        \toprule
        \textbf{Metrics} & \textbf{Description} & \textbf{Calculation} & \textbf{Unit} \\
        \midrule

        Travel time $t$
        & Total time of a vehicle traversing the segment
        & $t_C - t_A$
        & s \\

        Travel speed $v$
        & Average travel speed along the segment
        & $\lvert S_C - S_A \rvert / \lvert t_C - t_A \rvert$
        & mph \\

        Free-flow speed $v_f$
        & Desired vehicle speed unaffected by background traffic
        & 80$^{\mathrm{th}}$ percentile of speeds in free-flow states
        & mph \\

        Control delay $d_c$
        & Vehicle delay caused by signal control
        & $t_C - t_B$
        & s \\

        Level of Service
        & Level of service ranging from A to F
        & Determined by $t_C$
        & A--F \\

        Stop delay $d_s$
        & Total stop duration within the segment
        & $\sum_{i=1}^{n_s} \lvert t_{i,e} - t_{i,s} \rvert$
        & s \\

        Number of stops $n_s$
        & Number of stops within the intersection
        & Number of stop states
        & -- \\

        Queue length $q$
        & Maximum distance to the stop bar among all stop states
        & $\lvert S_D - S_C \rvert$
        & m \\

        Arrive on green (AOG)
        & Whether a vehicle arrives during the green phase
        & Signal state at Point B or $n_s \ge 1$
        & $\{0,1\}$ \\

        Split failure (SF)
        & Vehicle fails to pass the intersection within one cycle
        & Determined by $n_s \ge 2$
        & $\{0,1\}$ \\

        \bottomrule
    \end{tabularx}
\end{table}

\textbf{Travel time $t$} \& \textbf{Travel speed $v$.}
Travel time is defined as the total time for a vehicle to traverse an arterial link, and travel speed is defined as the total travel distance divided by travel time. As shown in \cref{fig:4}, they are computed as
\begin{equation}
t = t_C - t_A, 
\qquad
v = L/t = \lvert S_C - S_A \rvert / \lvert t_C - t_A \rvert.
\label{eq:tt_v}
\end{equation}

\textbf{Free-flow speed $v_f$} \& \textbf{Free-flow arrival time $t_B$.}
Free-flow speed is defined as the desired speed when a vehicle is not affected by other vehicles. Due to heterogeneous driving behaviors, free-flow speed may vary across vehicles.
Following prior work \citep{Wang2023Trajectory}, $v_f$ is estimated as the 80$^{\mathrm{th}}$ percentile of the speeds during all free-flow states; if no free-flow state is detected, $v_f$ is set to the link speed limit. Given $v_f$, the free-flow arrival time is computed as
\begin{equation}
t_B = (S_B - S_A)/v_f.
\label{eq:tf}
\end{equation}

\textbf{Control delay $d_c$} \& \textbf{Level of Service (LOS).}
Control delay is defined as the difference between the actual travel time and the free-flow travel time. 
With the estimated free-flow arrival time $t_B$, control delay is calculated as
\begin{equation}
d_c = t_C - t_B,
\label{eq:dc}
\end{equation}
Based on the resulting delay, \citet{HCM2022} provides a level-of-service rating from A to F.

\textbf{Stop delay $d_s$} \& \textbf{Number of stops $n_s$.}
Based on the stop states obtained from trajectory segmentation, the number of stops can be counted as $n_s$. 
Total stop delay is defined as the sum of the durations of all stop states (see \cref{fig:4}):
\begin{equation}
d_s = \sum_{i=1}^{n_s} \lvert t_{i,e} - t_{i,s} \rvert,
\label{eq:ds}
\end{equation}
where $t_{i,s}$ and $t_{i,e}$ denote the start and end times of the $i$-th stop state, respectively.

\textbf{Queue length $q$.}
Considering multiple stop cases, queue length is defined as the distance from the stop bar to the first stop state, consistent with prior studies \citep{SaldivarCarranza2023Next,Wang2023Trajectory,Wang2024aTraffic}:
\begin{equation}
q = \lvert S_D - S_C \rvert.
\label{eq:q}
\end{equation}

\textbf{Arrival on green (AOG)} \& \textbf{Split failure (SF).}
A vehicle is labeled as AOG if it arrives during the green phase when traveling at free-flow speed. 
Since signal timing is unavailable for city-scale applications, AOG is approximated as $\mathrm{AOG}=0$ if the vehicle stops at least once (i.e., $n_s \ge 1$). 
Split failure occurs when a vehicle fails to pass the intersection within one signal cycle; accordingly, a trajectory is labeled as split failure ($\mathrm{SF}=1$) if it stops more than once (i.e., $n_s \ge 2$).

\subsubsection{Link-level traffic state measures}
Link-level traffic state measures are obtained by aggregating single-vehicle traffic state measures within specific time windows. 
Specifically, for a given link $\mathcal{L}_i$, if $N$ vehicles traverse the link during a time window $T$, the aggregated link-level measures are computed as:
\begin{equation}
\begin{alignedat}{4}
T_T^{\mathcal{L}_i}   &= \tfrac{1}{N}\sum_{k=1}^{N} t_k, \quad &
V_T^{\mathcal{L}_i}   &= \tfrac{1}{N}\sum_{k=1}^{N} v_k, \quad &
CD_T^{\mathcal{L}_i}  &= \tfrac{1}{N}\sum_{k=1}^{N} d_{c,k}, \quad &
SD_T^{\mathcal{L}_i}  &= \tfrac{1}{N}\sum_{k=1}^{N} d_{s,k}, \\
S_T^{\mathcal{L}_i}   &= \tfrac{1}{N}\sum_{k=1}^{N} n_{s,k}, \quad &
Q_T^{\mathcal{L}_i}   &= \tfrac{1}{N}\sum_{k=1}^{N} q_k, \quad &
AOG_T^{\mathcal{L}_i} &= \tfrac{1}{N}\sum_{k=1}^{N} \mathrm{AOG}_k, \quad &
SF_T^{\mathcal{L}_i}  &= \tfrac{1}{N}\sum_{k=1}^{N} \mathrm{SF}_k .
\end{alignedat}
\end{equation}

\begin{figure}
    \centering
    \includegraphics[width=0.75\textwidth]{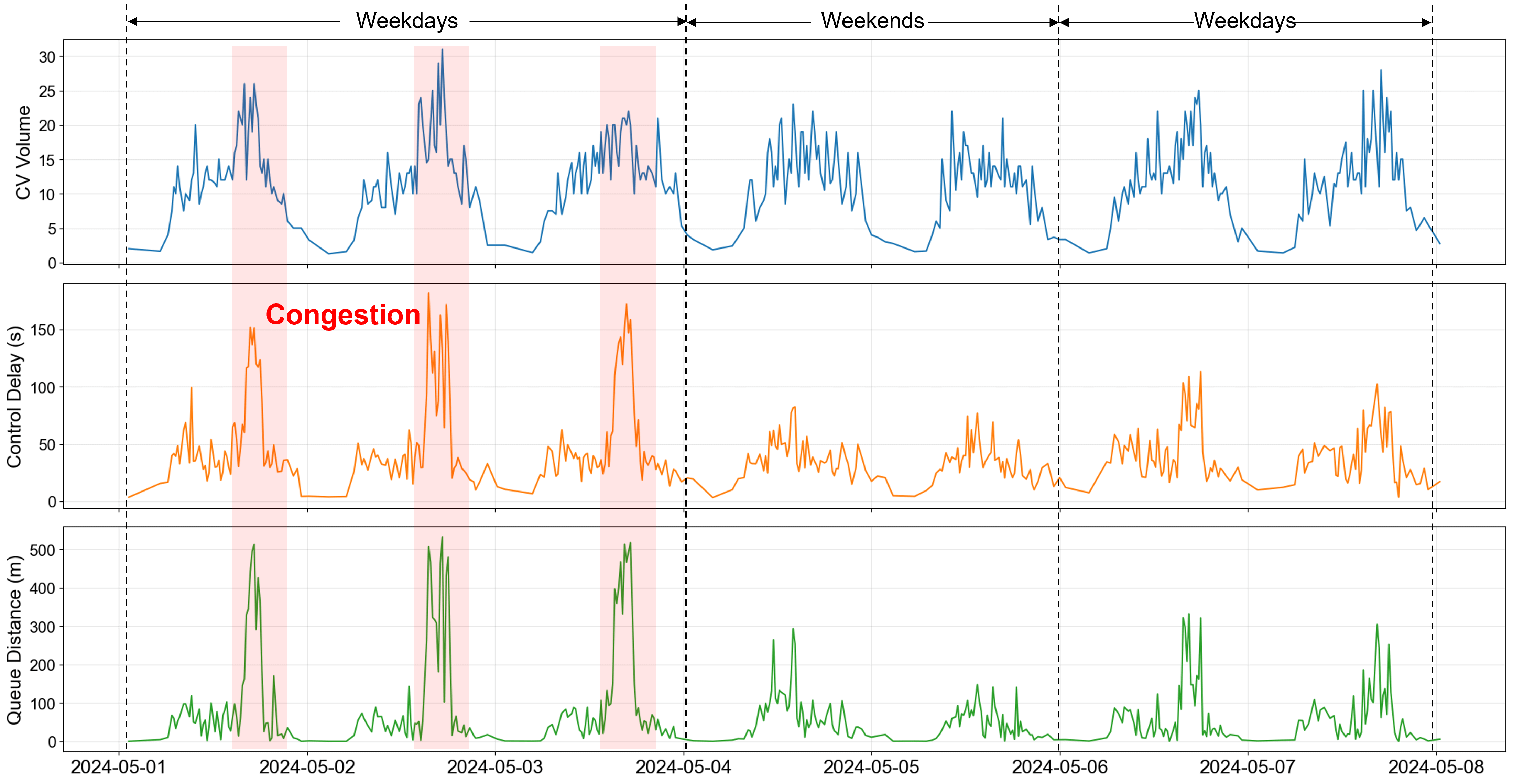}
    \vspace{-1em} 
    \caption{Example of link-level traffic state measures within one week.}
    \label{fig:5}
\end{figure}

Given the low CV penetration rate in current practices (e.g., 2--3\% in this study), it is hard to aggregate these measures over a short time interval (e.g., 5\,min or cycle-level) as there will be substantial missing observations due to insufficient samples. 
By examining the empirical distribution of CV counts on arterials, we select a 15\,min time window to ensure a sufficient number of CV samples ($N\ge 10$) in most cases (90.7\%) while retaining a fine temporal granularity, consistent with prior studies \citep{Novak2023DataDriven,Zhao2020TGCN}. 
Therefore, the aggregated measures represent the average traffic operating states within each time window. Under the assumption of uniformly random CV arrivals, the maximum queue length is theoretically expected to be twice the observed average queue length \citep{Tavafoghi2021Queue}, supporting the use of average queue length as a reliable proxy for overall queue conditions.
\Cref{fig:5} visualizes key traffic state measures for a representative arterial link over one week. A pronounced weekday periodicity is observed: weekday traffic typically exhibits higher volumes, larger delays, and longer queues than weekends. 
Congestion during the first three days is reflected by substantially elevated control delays and queue lengths.

\section{Abnormality-Aware Arterial Traffic Prediction}
\label{sec:aastgcn}
\subsection{Abnormal Traffic Identification}
\begin{figure}
    \centering
    \includegraphics[width=0.85\textwidth]{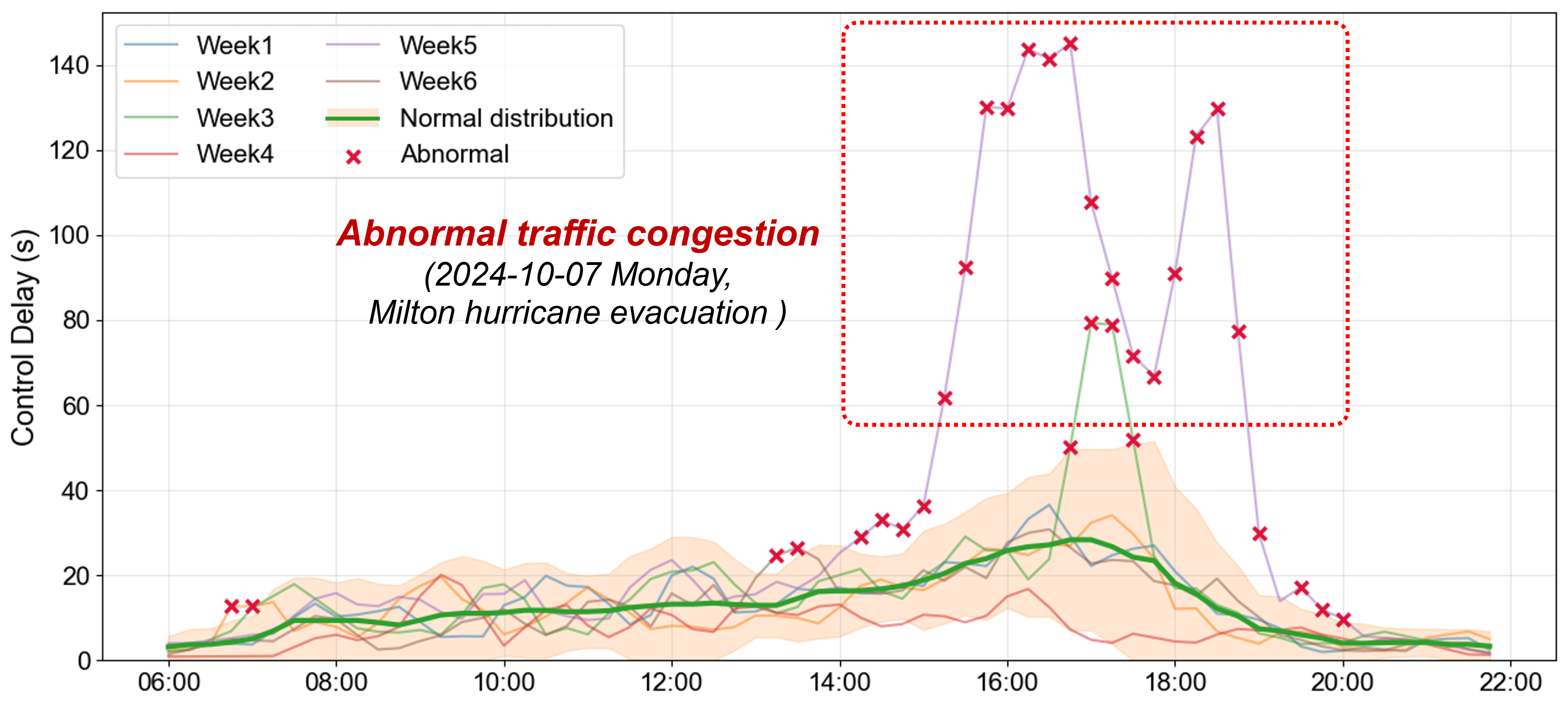}
    \vspace{-1em} 
    \caption{Illustration of abnormal traffic identification.}
    \label{fig:6}
\end{figure}

Abnormal traffic may arise from both planned events (e.g., sporting games) and unplanned incidents (e.g., crashes), causing traffic conditions that differ substantially from typical patterns. 
Existing studies often assume that such event information is known \emph{a priori} before prediction \citep{Essien2021Deep}. 
In practical arterial management, however, this information is frequently unavailable in real time, especially for unplanned events. 
To address this issue, we adopt a data-driven, median-based method to identify abnormal traffic by comparing real-time observations with historical traffic patterns, as shown in \cref{fig:6}.
For a given link at time $t$, we construct a historical reference set $W_t$ consisting of observations at the same time on the same day of week, i.e., $\tau \in W_t$. 
Let $x_t$ denote a traffic measure (e.g., delay), its historical median $m_t$ and a scale estimate $\sigma_t$ based on the median absolute deviation (MAD) are computed as
\begin{equation}
m_t = \mathrm{Median}\left(\{x_\tau\}_{\tau \in W_t}\right),
\qquad 
\mathrm{MAD}_t = \mathrm{Median}\left(\{\lvert x_\tau - m_t \rvert\}_{\tau \in W_t}\right), 
\qquad 
\sigma_t = 1.4826 \cdot \max\!\left(\mathrm{MAD}_t, \epsilon\right)
\label{eq:mad_sigma}
\end{equation}
An abnormal condition is flagged when $x_t$ exceeds an upper threshold:
\begin{equation}
\tilde{x}_t = m_t + k\sigma_t,
\qquad
\mathrm{Abnormal}(x_t)=
\begin{cases}
1, & x_t \ge \tilde{x}_t,\\
0, & \text{otherwise}.
\end{cases}
\label{eq:abnormal_rule}
\end{equation}
Here, the constant 1.4826 scales MAD to be consistent with the Gaussian standard deviation, $\epsilon$ avoids degenerate cases, and $k$ controls the sensitivity of detection. 
Our experiment found that compared with mean-based rules, this Median-based rule is more robust to outliers in limited-period CV data, which is consistent with existing studies \citep{Leys2013Detecting}. 
As an example, abnormal congestion caused by hurricane evacuation is successfully detected in \cref{fig:6}. This distribution-aware rule does not require \emph{a priori} knowledge of abnormal events and can be applied in real time using only historical traffic data.

\subsection{Abnormality-Aware Spatiotemporal Graph Convolution Network (AASTGCN)}
\subsubsection{Preliminaries}

\begin{figure}
    \centering
    \includegraphics[width=0.85\textwidth]{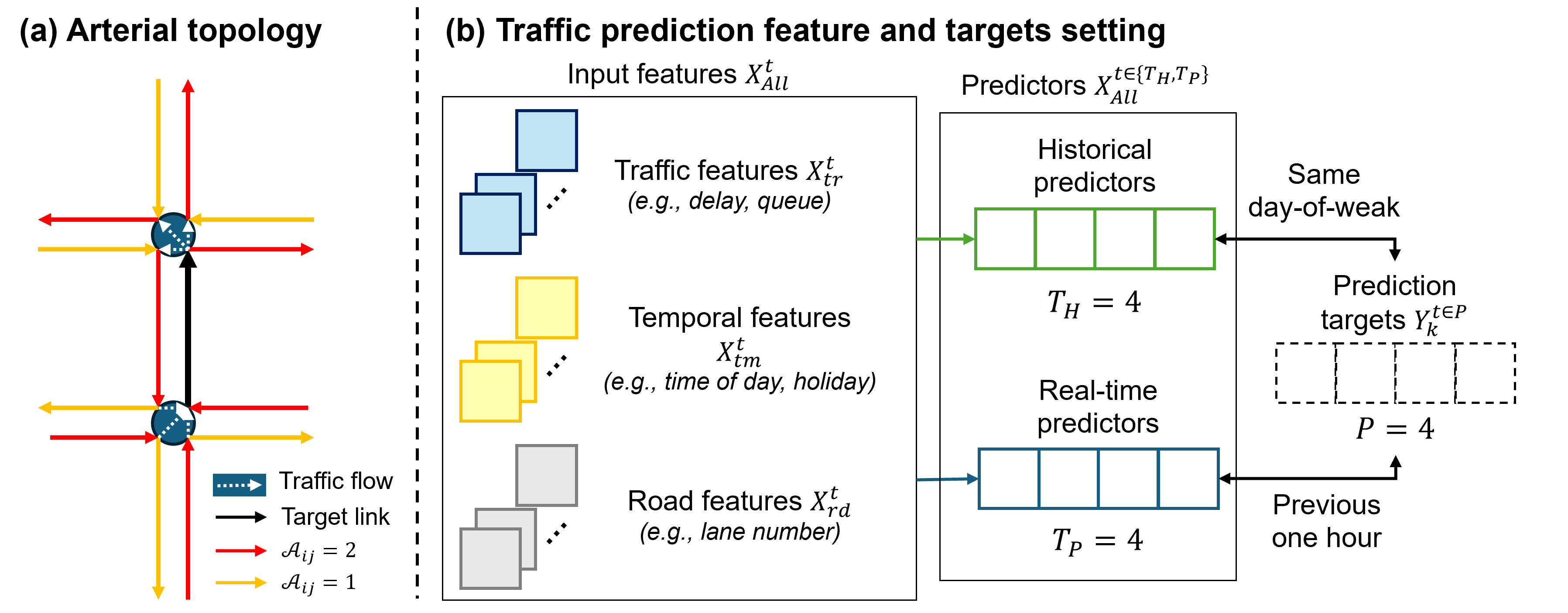}
    \vspace{-1em} 
    \caption{Illustration of traffic state prediction preliminaries.}
    \label{fig:7}
\end{figure}

\textbf{Arterial network.}
Given an arterial network $\{\mathcal{L}_i\}_{i=1}^{N}$, a fixed adjacency
matrix $\mathbf{A}_{\mathrm{fix}} \in \mathbb{R}^{N\times N}$ is constructed based on
link-to-link topology, as shown in \cref{fig:7}(a). For any pair of links $\mathcal{L}_i$
and $\mathcal{L}_j$, the adjacency entry is defined as
\begin{equation}
(\mathbf{A}_{\mathrm{fix}})_{ij} =
\begin{cases}
2, & \text{if } \mathcal{L}_i \text{ and } \mathcal{L}_j \text{ share the same intersection and are connected by direct movements},\\
1, & \text{if } \mathcal{L}_i \text{ and } \mathcal{L}_j \text{ share the same intersection but have no direct traffic connection},\\
0, & \text{otherwise}.
\end{cases}
\label{eq:afix_def}
\end{equation}
The resulting $\mathbf{A}_{\mathrm{fix}}$ is then normalized into a row-stochastic adjacency
matrix to represent the network structure.

\textbf{Input features.}
Given a link $\mathcal{L}_i$ at time step $t$, the feature vector
$\mathbf{X}_{\mathrm{All}}^{(t,\mathcal{L}_i)}$ consists of three categories, as shown in \cref{fig:7}(b). 
Traffic features
$\mathbf{X}_{\mathrm{tr}}^{(t,\mathcal{L}_i)}$ represent dynamic traffic states including average control delay, queue length, speed, CV volume, travel time, and AOG ratio.
Temporal features 
$\mathbf{X}_{\mathrm{tm}}^{(t,\mathcal{L}_i)}$ capture periodic time
indicators including time-of-day, day-of-week, a holiday flag, and peak-hour indicators (morning peak, afternoon peak, and off-peak). 
Road features
$\mathbf{X}_{\mathrm{rd}}^{(t,\mathcal{L}_i)}$ include static roadway attributes such as road ID, lane number, and speed limit.

\textbf{Problem statement: traffic prediction.}
Given the arterial network structure and historical input features, the objective is to predict future one-hour traffic states. Let $Y_k^{(t\in \mathcal{P})}$ denote the future target for metric $k$ over the prediction horizon $\mathcal{P}$, where $k$ corresponds to link control delay and queue length in this study. Let
$\mathbf{X}_{\mathrm{All}}^{t}$ denote the features for all links at time $t$.
Two temporal windows are used as predictors. The real-time predictors consist of a sequence of the past $T_P=4$ time steps (i.e., the previous one hour) to capture short-term traffic dynamics. The historical predictors are constructed as a historical average sequence of $T_H=4$ time steps from the same prediction period on the same day of week to capture long-term periodicity. The prediction task is to learn a mapping function $f(\cdot)$ such that
\begin{equation}
\hat{\mathbf{Y}}_{k}^{(t+1,\ldots,t+P)} =
f\!\left(
\{\mathbf{X}_{\mathrm{All}}^{(t-T_P+1)},\ldots,\mathbf{X}_{\mathrm{All}}^{t}\},
\{\mathbf{X}_{\mathrm{All}}^{(t_{\mathrm{hist}}+1)},\ldots,\mathbf{X}_{\mathrm{All}}^{(t_{\mathrm{hist}}+P)}\},
\mathbf{A}_{\mathrm{fix}}
\right).
\label{eq:pred_mapping}
\end{equation}

\subsubsection{AASTGCN framework overview}
\begin{figure}
    \centering
    \includegraphics[width=0.95\textwidth]{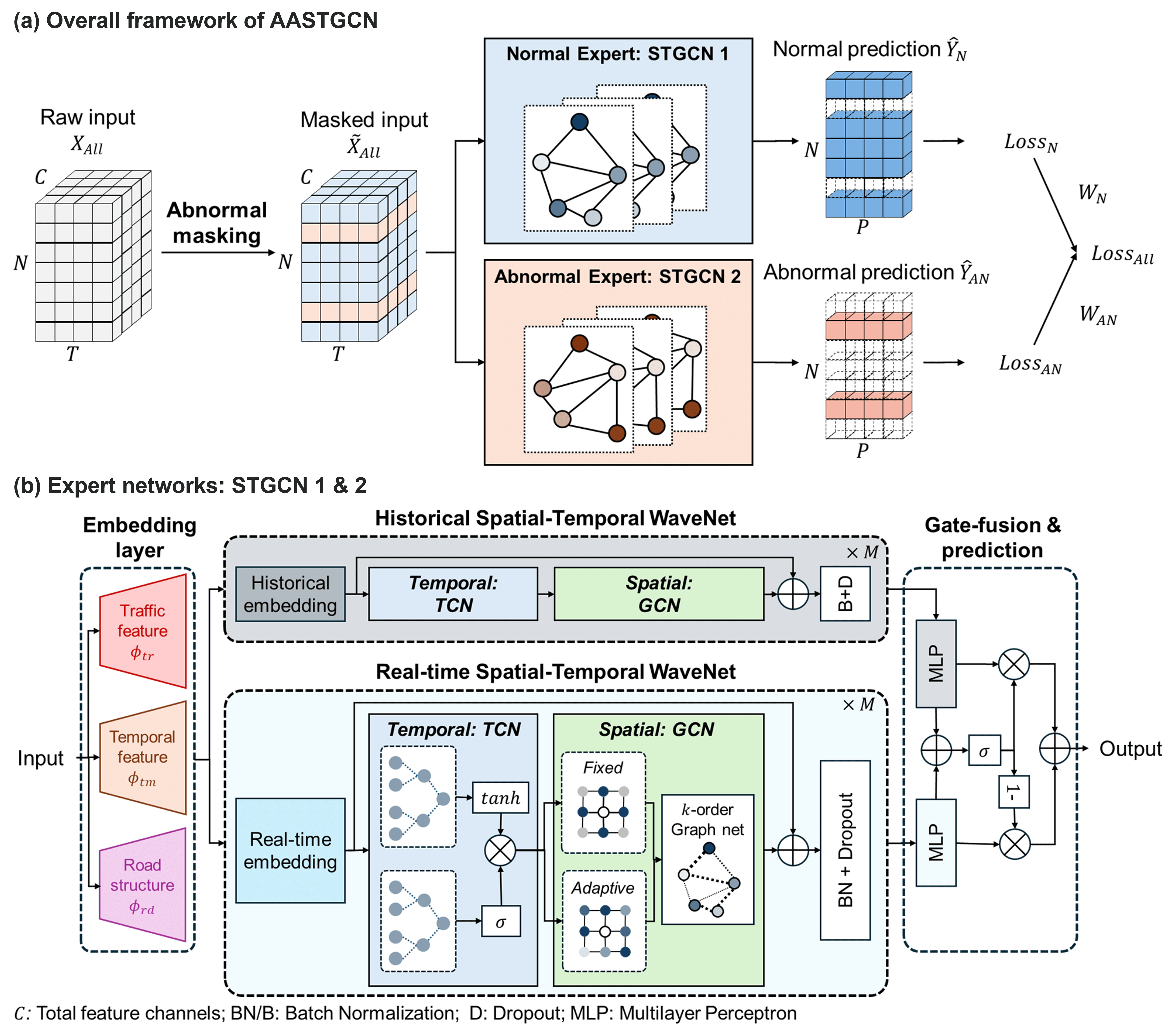}
    \vspace{-1em} 
    \caption{AASTGCN framework and expert network structure.}
    \label{fig:8}
\end{figure}

\cref{fig:8}(a) illustrates the overall framework of the AASTGCN. 
Given the raw inputs $\mathbf{X}_{\mathrm{All}}$, an abnormal masking is first conducted at the current time step $t$. 
For example, when predicting traffic states for the period from 8:00 to 9:00\,a.m., the abnormal mask is determined based on the link-level abnormal labels from the most recent observed interval (i.e., 7:45--8:00\,a.m.). 
The masked inputs $\tilde{\mathbf{X}}_{\mathrm{All}}$ are then fed into two expert networks, namely the normal expert and the abnormal expert, to generate the corresponding predictions. The outputs from the two experts, denoted as $\hat{\mathbf{Y}}_{\mathrm{N}}$ and $\hat{\mathbf{Y}}_{\mathrm{AN}}$, are used to compute their respective losses, $\mathcal{L}_{\mathrm{N}}$ and $\mathcal{L}_{\mathrm{AN}}$. 
These losses are subsequently combined through a weighted aggregation to form the overall training objective $\mathcal{L}_{\mathrm{All}}$, which jointly optimizes both experts. This abnormality-aware framework enables the two experts to learn distinct traffic patterns without being confounded by mixed training, thereby improving prediction robustness under both normal and abnormal traffic conditions.

\subsubsection{Expert networks}
As shown in \cref{fig:8}(b), the two experts adopt the same STGCN network structure. Specifically, the raw inputs are first passed through an embedding layer to obtain a rich latent representation. Next, the historical and real-time embeddings are encoded by two dedicated spatiotemporal WaveNet modules to capture their distinct spatiotemporal dependencies. Finally, a gate-based feature fusion module followed by a prediction head integrates the historical and real-time information to generate the final outputs.

\textbf{Embedding layer.}
To effectively integrate heterogeneous traffic data, an embedding layer is first applied to map the raw input features into a unified latent representation. A hybrid encoding strategy with three parallel MLP branches is adopted to embed continuous traffic variables, categorical temporal attributes, and static road information:
\begin{equation}
\mathbf{E}_{\mathrm{All}}
=
\big[
\phi_{\mathrm{tr}}(\mathbf{X}_{\mathrm{tr}})
\;\oplus\;
\phi_{\mathrm{tm}}(\mathbf{X}_{\mathrm{tm}})
\;\oplus\;
\phi_{\mathrm{rd}}(\mathbf{X}_{\mathrm{rd}})
\big],
\label{eq:embedding}
\end{equation}
where $\oplus$ denotes the concatenation operation. The function $\phi_{\mathrm{tr}}(\cdot)$
represents a linear projection that maps continuous traffic features (e.g., control delay)
into a high-dimensional latent space. For categorical temporal and road features (e.g., time
of day and speed limit), their discrete indices are first encoded and then mapped into dense
embedding vectors via $\phi_{\mathrm{tm}}(\cdot)$ and $\phi_{\mathrm{rd}}(\cdot)$ to capture their semantic properties. The final embedded representation $\mathbf{E}_{\mathrm{All}} \in \mathbb{R}^{N \times T \times D}$ is obtained to jointly exploit dynamic traffic states, temporal contexts, and the underlying road structure.

\textbf{Spatial--temporal WaveNet.}
Following existing studies \citep{Li2024STFGCN,Zou2026MultiGraph}, a spatiotemporal (ST) WaveNet is developed
to capture the spatiotemporal dependencies in traffic data. The same ST-WaveNet architecture
is applied to both the real-time embedding $\mathbf{E}_{\mathrm{All}}^{(t \in \mathcal{T}_P)}$
and the historical embedding $\mathbf{E}_{\mathrm{All}}^{(t \in \mathcal{T}_H)}$, enabling the
model to learn congruent spatiotemporal features from short-term traffic dynamics and
long-term periodic patterns. Within each ST-WaveNet, a Temporal Convolutional Network (TCN)
is first used to model temporal correlations, followed by an adaptive Graph Convolutional Network (GCN) to capture spatial dependencies among arterial links. Residual and skip connections are further incorporated to ensure training stability and multi-scale feature aggregation.

\textit{Temporal convolution network (TCN).}
The core component of the TCN is the Dilated Causal Convolution (DCC), which has demonstrated
superior performance over traditional RNN-based models in time-series prediction tasks. For
a one-dimensional sequence $\mathbf{X} \in \mathbb{R}^{T}$ with a filter
$\mathbf{f} \in \mathbb{R}^{K}$, the dilated causal convolution at time step $t$ is defined
as
\begin{equation}
(\mathbf{X} * \mathbf{f})(t)
=
\sum_{i=0}^{K-1} \mathbf{f}(i)\,\mathbf{X}(t - d i),
\label{eq:dcc}
\end{equation}
where $d$ denotes the dilation factor controlling the temporal skipping distance. Stacking
DCC layers with exponentially increasing dilation factors allows the receptive field to
expand exponentially, enabling the model to capture long-range temporal dependencies using
a limited number of layers. A Gated Linear Unit (GLU) mechanism is then adopted to regulate
the information flow between two parallel TCNs. Given an input
$\mathbf{S} \in \mathbb{R}^{N \times T}$, the GLU is formulated as
\begin{equation}
\mathbf{h}
=
g(\mathbf{W}_1 (\mathbf{S} * \mathbf{f}) + \mathbf{b})
\;\otimes\;
\sigma(\mathbf{W}_2 (\mathbf{S} * \mathbf{f}) + \mathbf{c}),
\label{eq:glu}
\end{equation}
where $\mathbf{W}_1$, $\mathbf{W}_2$, $\mathbf{b}$, and $\mathbf{c}$ are learnable parameters,
$\otimes$ denotes the element-wise product, $g(\cdot)$ is an activation function (tanh in this study), and $\sigma(\cdot)$ is the sigmoid function controlling the information gating ratio.

\textit{Adaptive GCN.}
Following each TCN layer, an adaptive GCN is integrated to model spatial correlations among arterial links. As the predefined topology-based adjacency matrix
$\mathbf{A}_{\mathrm{fix}}$ alone cannot fully capture complex and time-varying spatial dependencies in traffic systems, a data-driven adaptive
adjacency matrix $\mathbf{A}_{\mathrm{apt}}$ is learned as
\begin{equation}
\mathbf{A}_{\mathrm{apt}}
=
\mathrm{softmax}\!\left(\mathrm{ReLU}(\mathbf{E}_1 \mathbf{E}_2^{\top})\right),
\label{eq:adaptive_adj}
\end{equation}
where $\mathbf{E}_1, \mathbf{E}_2 \in \mathbb{R}^{N \times d}$ are learnable node embeddings
for source and target nodes, respectively. The Singular Value Decomposition (SVD) of
$\mathbf{A}_{\mathrm{fix}}$ is used to initialize $\mathbf{E}_1$ and $\mathbf{E}_2$, allowing
the adaptive adjacency to refine the spatial structure from a physically informed baseline.
With both $\mathbf{A}_{\mathrm{fix}}$ and $\mathbf{A}_{\mathrm{apt}}$, the $k$-order graph convolution is formulated:
\begin{equation}
\mathbf{z}_{\mathrm{out}}
=
\sum_{i=0}^{k}
\left(
\mathbf{h}\mathbf{A}_{\mathrm{fix}}^{i}\mathbf{W}_{i,1}
+
\mathbf{h}\mathbf{A}_{\mathrm{apt}}^{i}\mathbf{W}_{i,2}
\right),
\label{eq:gcn}
\end{equation}
where $k$ denotes the diffusion step and $\mathbf{W}_{i,1}$ and $\mathbf{W}_{i,2}$ are
learnable parameters.

\textit{Residual and skip connections.}
To stabilize training, a residual connection is employed by adding the layer input
$\mathbf{X}_{\mathrm{in}}$ to the output of the GCN,
$\mathbf{X}_{\mathrm{res}} = \mathbf{z}_{\mathrm{out}} + \mathbf{X}_{\mathrm{in}}$.
The outputs from all hidden layers are further projected through $1\times1$ convolutions
and aggregated via skip connections
$\mathbf{H}_{\mathrm{skip}} = \sum \mathrm{Conv}(\mathbf{z}_{\mathrm{out}})$.
A ReLU activation is finally applied to obtain the fused spatiotemporal representation.

\textbf{Gate fusion and prediction.}
To integrate short-term traffic fluctuations with long-term periodic patterns, a gate-based
fusion module followed by a prediction head is designed, as illustrated in \cref{fig:8}(b). The
output representations from the real-time ST-WaveNet $\mathbf{H}_{\mathrm{rt}}$ and the
historical ST-WaveNet $\mathbf{H}_{\mathrm{hist}}$ are first transformed by corresponding
MLPs. A gated fusion mechanism is then applied:
\begin{equation}
\mathbf{g}
=
\sigma\!\left(
\mathbf{W}_g
\big[
\mathbf{H}_{\mathrm{rt}} \oplus \mathbf{H}_{\mathrm{hist}}
\big]
+
\mathbf{b}_g
\right),
\qquad
\mathbf{H}_{\mathrm{fusion}}
=
\mathbf{g} \otimes \mathrm{MLP}_{\mathrm{rt}}(\mathbf{H}_{\mathrm{rt}})
+
(1-\mathbf{g}) \otimes \mathrm{MLP}_{\mathrm{hist}}(\mathbf{H}_{\mathrm{hist}}),
\label{eq:fusion}
\end{equation}
where $\mathbf{g} \in (0,1)^D$ is a learnable gating vector whose elements are constrained to
the interval $(0,1)$. This mechanism allows the model to adaptively balance real-time traffic
dynamics and historical traffic patterns, prioritizing real-time information during
abnormal events (e.g., crashes) and historical trends during stable, recurrent periods.

The fused representation $\mathbf{H}_{\mathrm{fusion}}$ is finally passed through an MLP
followed by a ReLU activation to generate non-negative traffic state predictions:
\begin{equation}
\hat{Y}_k^{(t+1,\ldots,t+P)}
=
\mathrm{ReLU}\!\left(
\mathrm{MLP}(\mathbf{H}_{\mathrm{fusion}})
\right).
\label{eq:prediction}
\end{equation}

\subsubsection{Loss function}
Finally, the Smooth~L1 loss is adopted as the loss function for both the normal and abnormal experts due to its robustness to extreme values and effectiveness in preventing gradient explosion. The loss for each expert is defined as
\begin{equation}
Loss_i
=
\frac{1}{P N}
\sum_{p=1}^{P}
\sum_{n=1}^{N}
\mathrm{Smooth}_{\mathrm{L1}}
\!\left(
Y_k^{(p,n)} - \hat{Y}_{k,i}^{(p,n)}
\right),
\label{eq:loss_expert}
\end{equation}
where $i \in \{\mathrm{N}, \mathrm{AN}\}$ denotes the normal and abnormal experts, respectively.
Here, $P$ is the prediction horizon and $N$ is the number of links. The final training objective is formulated as a weighted combination of the two expert losses:
\begin{equation}
Loss_{\mathrm{All}}
=
W_{\mathrm{N}}*Loss_{\mathrm{N}}
+
W_{\mathrm{AN}}*Loss_{\mathrm{AN}},
\label{eq:loss_total}
\end{equation}
where $W_{\mathrm{N}}$ and $W_{\mathrm{AN}}$ are the loss weights for the normal and abnormal
experts, respectively, and their sum is constrained to unity. Since abnormal cases are
relatively rare, $W_{\mathrm{AN}}$ is set to be larger than $0.5$ (e.g., $0.7$--$0.8$) to
penalize prediction errors from the abnormal expert more heavily. This design ensures that
the model allocates sufficient learning capacity to rare yet critical abnormal traffic
conditions.

\section{Experiments}
\label{sec:experiments}
\subsection{Data Description}
\begin{figure}
    \centering
    \includegraphics[width=0.8\textwidth]{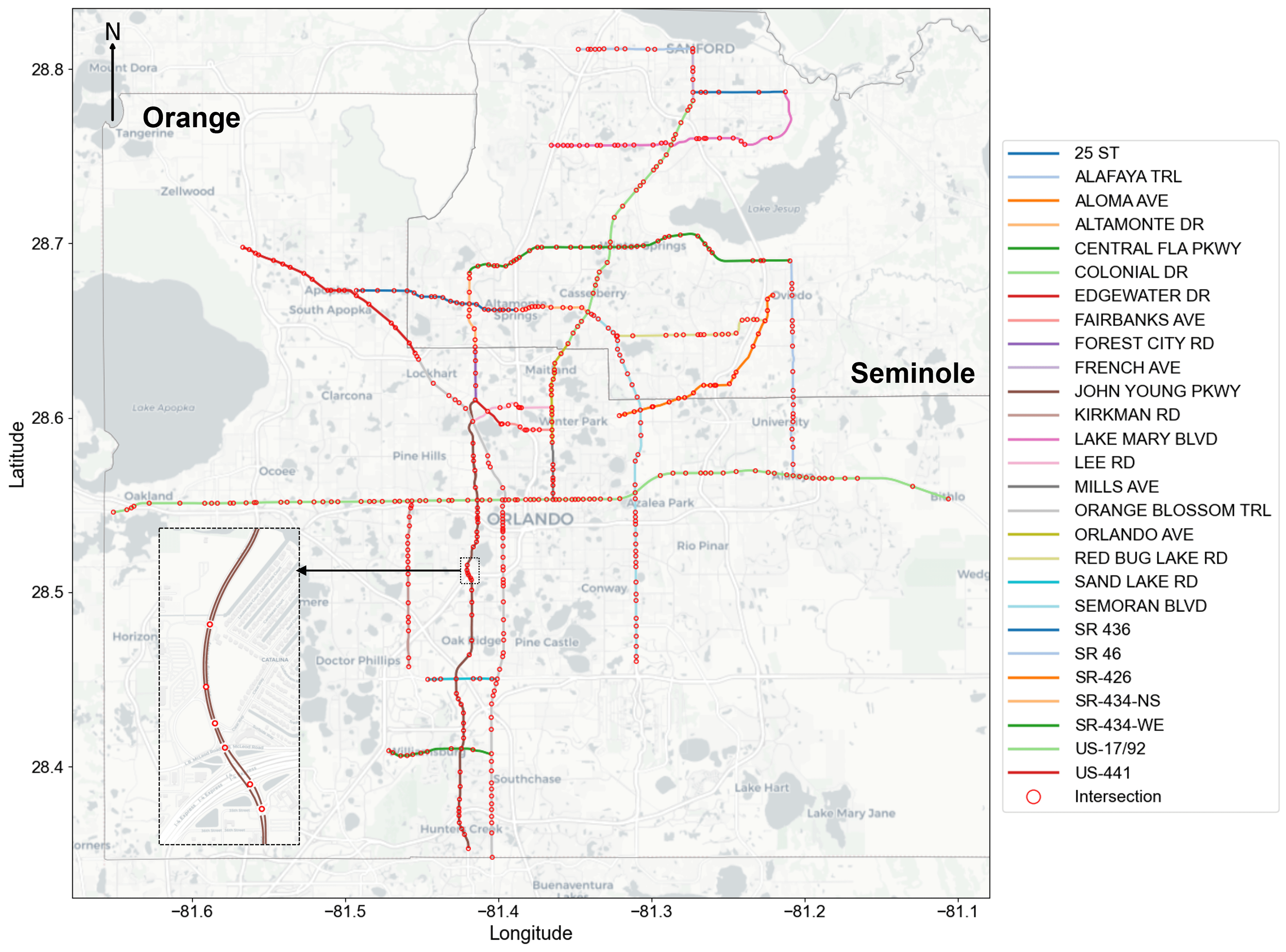}
    \vspace{-1em} 
    \caption{City-scale arterial networks.}
    \label{fig:9}
\end{figure}

In this study, we use StreetLight CV data from two Florida counties: Orange and Seminole. This dataset covers 45 days split between May 1-31 and Oct 1-15, 2024. 
Specifically, it contains 3-s interval vehicle trajectories collected from original equipment manufacturers. 
Each CV point includes journey Id, timestamp, GPS location (latitude, longitude), heading, and speed information. 
On average, it comprises over 8.9 million CV trajectory points from 404,997 journeys per day. 
As shown in \cref{fig:9}, a total of 27 major arterials (e.g., Colonial DR, Alafaya TRL) are selected and extend over 386.4 miles (including both separate directions). 
After road segmentation, 1,050 directional links are obtained, forming a city-scale arterial network that is much larger than commonly used datasets (e.g., 627 links in DiDi-SZ and 307-883 nodes in PeMS datasets). 
Detailed link information is summarized in \cref{tab:additional_results}. 
By comparing the CV volume with 2024 AADT, the estimated CV penetration rate in the study arterial networks ranges between about 2.2\%-3.6\%.

\begin{figure}
    \centering
    \includegraphics[width=0.85\textwidth]{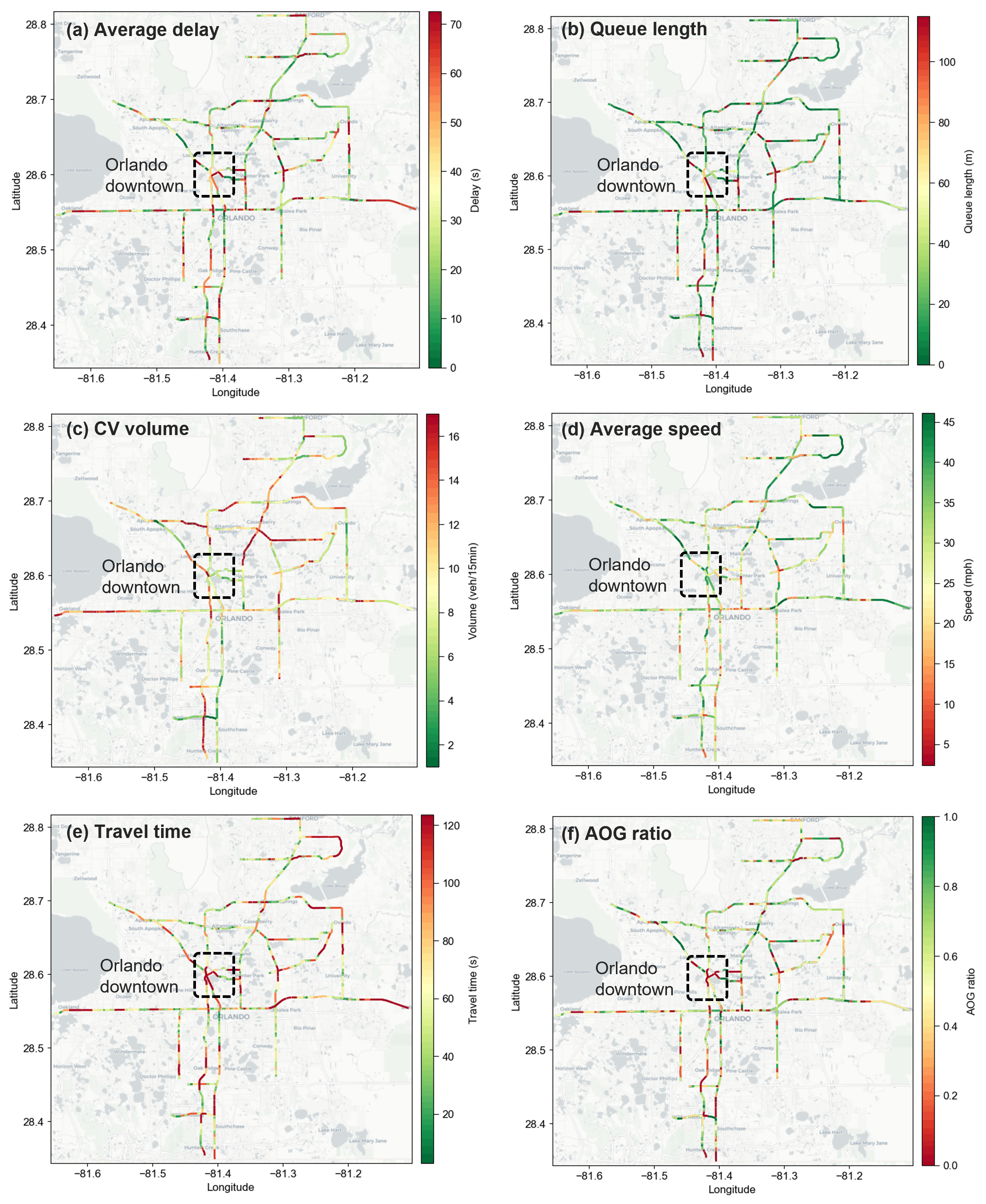}
    \vspace{-1em} 
    \caption{Network-level traffic state measures (2024-05-01 8:30-8:45).}
    \label{fig:10}
\end{figure}

Based on the proposed traffic state extraction method, network-level traffic state measures can be estimated. 
Taking a Wednesday morning peak (08:30–08:45) on May 1st, 2024 as an example (\cref{fig:10}), network conditions exhibit pronounced spatial heterogeneity: congestion is concentrated on a subset of arterial links, while others operate under relatively light traffic. 
For example, arterial links in Orlando downtown area experience substantial congestion, characterized by relatively high average delay, queue length, CV volume, and travel time, along with a low AOG ratio. 

\subsection{Experiment Setup}
In the experiments, the processed network traffic measures are used as the modeling dataset. To mitigate the impact of low CV samples overnight, only data recorded between 06:00 and 22:00 each day are included. The train and test datasets are split by date to avoid potential information leakage: May 1-25 and Oct 1-7 (32 days) for training, and May 26-31 and Oct 11-15 (11 days) for testing. Data from Oct 8-10 are removed as these dates were affected by a hurricane and contain large missing recordings. The remaining missing values are imputed using temporal averaging. To find the best hyper-parameters of the proposed AASTGCN, a grid search is used and evaluated using 10\% of the train dataset as validation. 

To demonstrate the effectiveness of our proposed model, we also conducted traffic speed prediction experiments on the following prediction methods as baselines:

\textbf{LSTM}\textbf{: }A classical sequence model to capture temporal dependencies in traffic.

\textbf{GCN:} A classical graph model to capture spatial dependencies of traffic networks.

\textbf{DCRNN}\textbf{:} A diffusion convolutional recurrent neural network incorporating spatial and temporal dependency into traffic prediction \citep{Li2018Diffusion}.

\textbf{STGCN}\textbf{:} A spatio-temporal graph convolutional networks (GCN), which combines spatial graph convolution with 1-D temporal convolution \citep{Yu2018Spatio}.

\textbf{Graph-}\textbf{WaveNet}\textbf{: }A graph neural network that captures the spatial correlations with GCN layers and long sequences with dilated TCN \citep{Wu2019Graph}.

\textbf{ASTGCN}\textbf{:} An attention-based spatial–temporal GCN to model dynamic spatial–temporal correlations for traffic forecasting \citep{Guo2019Attention}.

\textbf{Conv-LSTM}\textbf{: }A spatiotemporal sequence-to-sequence model that integrates CNN and  LSTMs within an encoder–decoder framework \citep{Abdelraouf2022Utilizing}.

\textbf{iTransformer}\textbf{:} A novel transformer with inverted attention layers to effectively capture the complex traffic flow patterns and dependencies. \citep{Zou2024ShortTerm}.

\textbf{SATEformer}\textbf{:} A spatiotemporal attention-based Transformer that incorporates adaptive spatial and temporal embeddings \citep{Liu2023Spatio}.

\textbf{S}\textbf{TABC:} A spatiotemporal attention-based networks that incorporates attention-based GCN and dilated TCN modules \citep{Li2024STABC}.

In terms of the evaluation metrics, two commonly used metrics are adopted to quantify the differences between the observed values $Y_{i,j}$ and the predicted values $\hat{Y}_{i,j}$: the root mean square error (RMSE) and the mean absolute error (MAE):
\begin{equation}
    \label{eq:rmse_mae}
    \begin{aligned}
        \mathrm{RMSE}
        &=
        \sqrt{
        \frac{1}{DT}
        \sum_{i=1}^{D}
        \sum_{j=1}^{T}
        \left(
        Y_{i,j} - \hat{Y}_{i,j}
        \right)^2
        },
        \qquad
        \mathrm{MAE}
        &=
        \frac{1}{DT}
        \sum_{i=1}^{D}
        \sum_{j=1}^{T}
        \left|
        Y_{i,j} - \hat{Y}_{i,j}
        \right|.
    \end{aligned}
\end{equation}

where $D$ denotes the number of road segments and $T$ denotes the number of prediction time steps.

To ensure fair comparisons, we keep the same input features and use the same training and test sets for all baseline models. Each baseline was implemented according to its original specifications—either as described in the paper or in the official code—and then subjected to a grid search over its best model hyper-parameters. We selected the hyperparameter set that yields the highest validation performance for final testing. During model training, every model was run for up to 100 epochs with a batch size of 32. An early-stop mechanism was used in all experiments, and the number of early-stop epochs is set to 10, matching the AASTGCN’s settings. Furthermore, we implemented the AASTGCN and all baselines in PyTorch framework under the same hardware environment (NVIDIA RTX 3060 GPU), thereby ensuring consistency in running conditions.

\begin{table*}[t]
    \centering
    \caption{Overall performance comparison of different models.}
    \label{tab:4}
    \footnotesize\rmfamily
    \renewcommand{\arraystretch}{1.15}
    \setlength{\tabcolsep}{3pt}
    
    \begin{tabular}{lcccccccccccc}
        \toprule
        \multirow{3}{*}{\textbf{Models}}
        & \multicolumn{6}{c}{\textbf{Delay (s)}}
        & \multicolumn{6}{c}{\textbf{Queue (m)}} \\
        \cmidrule(lr){2-7}\cmidrule(lr){8-13}
        & \multicolumn{2}{c}{\makecell{\textbf{Overall} \\(n=604,800, 100\%)}}
        & \multicolumn{2}{c}{\makecell{\textbf{Normal} \\(n=589,037, 97.4\%)}}
        & \multicolumn{2}{c}{\makecell{\textbf{Abnormal} \\(n=15,763, 2.6\%)}}
        & \multicolumn{2}{c}{\makecell{\textbf{Overall} \\(n=604,800, 100\%)}}
        & \multicolumn{2}{c}{\makecell{\textbf{Normal} \\(n=577,848, 95.5\%)}}
        & \multicolumn{2}{c}{\makecell{\textbf{Abnormal} \\(n=26,952, 4.5\%)}}
        \\
        \cmidrule(lr){2-3}\cmidrule(lr){4-5}\cmidrule(lr){6-7}
        \cmidrule(lr){8-9}\cmidrule(lr){10-11}\cmidrule(lr){12-13}
        & MAE & RMSE & MAE & RMSE & MAE & RMSE
        & MAE & RMSE & MAE & RMSE & MAE & RMSE \\
        \midrule
        LSTM         & 7.559 & 14.383 & 7.089 & 12.446 & 23.988 & 44.979 & 10.866 & 27.497 & 9.894 & 23.448 & 30.089 & 69.540 \\
        GCN          & 5.718 & 10.550 & 5.402 &  9.176 & 16.749 & 32.521 &  7.754 & 18.827 & 7.121 & 16.192 & 20.267 & 46.679 \\
        DCRNN        & 4.817 & 10.230 & 4.562 &  8.897 & 13.738 & 31.559 &  6.654 & 20.543 & 6.120 & 17.500 & 17.219 & 52.068 \\
        STGCN        & 6.064 & 10.764 & 5.732 &  9.362 & 17.679 & 33.184 &  8.121 & 19.522 & 7.432 & 16.500 & 21.743 & 50.337 \\
        GraphWaveNet & 4.669 &  9.094 & 4.441 &  8.069 & 12.636 & 26.410 &  \underline{5.752} & \underline{15.472} & \underline{5.475} & 14.085 & 15.409 & 40.892 \\
        ASTGCN       & 5.051 & 10.922 & 4.739 &  9.288 & 15.966 & 35.683 &  6.930 & 19.874 & 6.338 & 16.771 & 18.638 & 51.419 \\
        Conv-LSTM    & 5.701 & 11.047 & 5.254 &  8.784 & 21.318 & 41.107 &  7.923 & 21.138 & 6.955 & 16.068 & 27.057 & 64.627 \\
        iTransformer &10.366 & 17.156 & 9.948 & 15.853 & 24.961 & 42.393 & 10.720 & 26.068 & 9.763 & 23.248 & 29.644 & 58.563 \\
        SATEformer   & 4.785 &  9.070 & 4.537 &  7.851 & 13.446 & 28.334 &  6.450 & 17.267 & 5.931 & 14.573 & 16.696 & 44.659 \\
        STABC        & \underline{4.423} &  \underline{8.660} & \underline{4.221} &  \underline{7.723} & \underline{11.500} & \underline{24.730} &  5.990 & 15.928 & 5.579 & \underline{13.934} & \underline{14.115} & \underline{37.828} \\
        \midrule
        \makecell[l]{\textbf{AASTGCN}\\\textit{(Improvement\textsuperscript{*})}} &
        \makecell{\textbf{3.535}\\\textit{(-20.1\%)}} &
        \makecell{\textbf{7.332}\\\textit{(-15.3\%)}} &
        \makecell{\textbf{3.328}\\\textit{(-21.2\%)}} &
        \makecell{\textbf{6.409}\\\textit{(-17.0\%)}} &
        \makecell{\textbf{10.747}\\\textit{(-6.6\%)}} &
        \makecell{\textbf{22.295}\\\textit{(-9.9\%)}} &
        \makecell{\textbf{4.819}\\\textit{(-15.6\%)}} &
        \makecell{\textbf{13.878}\\\textit{(-10.0\%)}} &
        \makecell{\textbf{4.436}\\\textit{(-18.6\%)}} &
        \makecell{\textbf{12.186}\\\textit{(-12.6\%)}} &
        \makecell{\textbf{12.401}\\\textit{(-12.2\%)}} &
        \makecell{\textbf{32.630}\\\textit{(-13.7\%)}} \\
        \bottomrule
    \end{tabular}
        
    \vspace{2pt}
    \begin{minipage}{0.98\linewidth}
        \footnotesize
        \emph{* Improvement is computed relative to the best-performing baseline model.}
    \end{minipage}
\end{table*}

\subsection{Experimental Results}
\subsubsection{Predicting performance comparison}
Based on the proposed abnormal detection method, 2.6\% of the samples are identified as abnormal cases for link-level average control delay (hereafter "delay"), while 4.5\% are detected as abnormal for link-level average queue length (hereafter "queue"), confirming that abnormal conditions are relatively rare among all traffic states. To provide a comprehensive evaluation, we report prediction errors not only for all test samples, but also separately for normal and abnormal subsets. 

\begin{figure}
    \centering
    \includegraphics[width=1\textwidth]{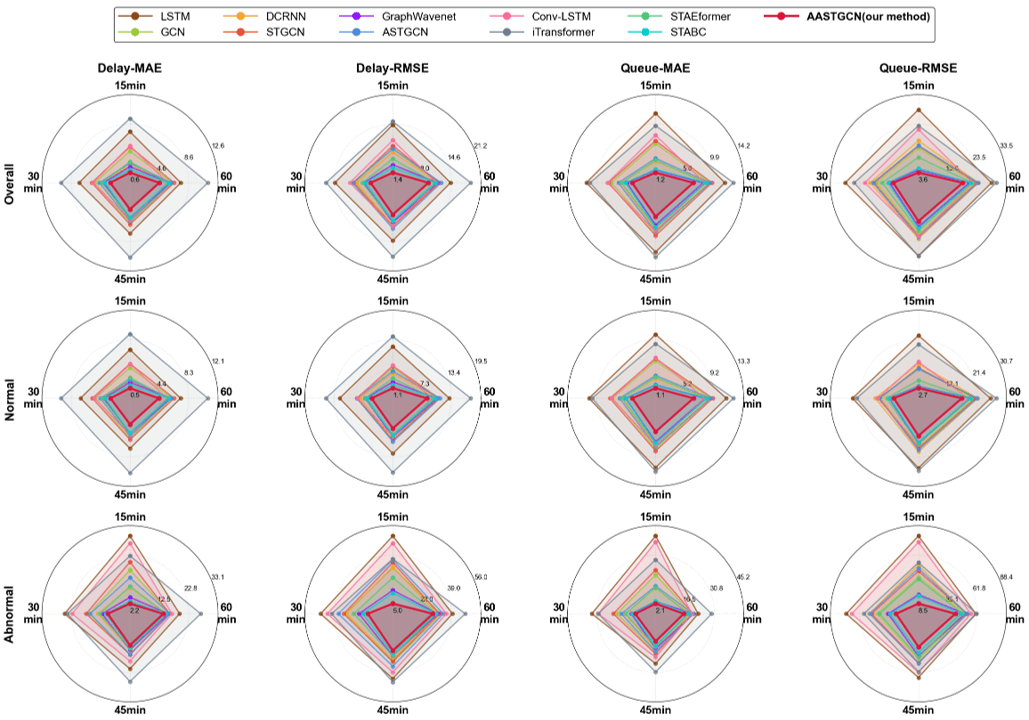}
    \vspace{-2em} 
    \caption{Model performance comparison at different prediction horizons (15, 30, 45, 60min).}
    \label{fig:11}
\end{figure}

\cref{tab:4} summarizes the prediction performance of the baseline models and the proposed AASTGCN on the test set. The best-performing method is highlighted in \textbf{bold}, and the runner-up is \underline{underlined}. Among the baselines, LSTM and iTransformer show the weakest performance because they only model temporal correlations. In contrast, models equipped with spatial modules (e.g., GCN-LSTM, DCRNN) achieve improved accuracy, indicating that incorporating spatial dependence is critical for arterial network traffic prediction. Among all baselines, Graph WaveNet, STAEformer, and STABC perform relatively well, benefiting from strong spatiotemporal modeling capabilities via graph convolutions and attention mechanisms. It is worth noting that prediction errors under abnormal conditions are substantially higher than those under normal conditions, indicating that abnormal traffic states are considerably more difficult to predict. This is because abnormal events often exhibit patterns that deviate markedly from regular traffic pattern and involve abrupt and irregular changes in traffic dynamics. The proposed AASTGCN consistently outperforms all baselines, achieving the lowest RMSE and MAE within both normal and abnormal cases. Unlike the baselines—which simply train normal and abnormal cases together—AASTGCN explicitly separates normal and abnormal traffic through abnormality-aware experts and leverages historical-periodic information, leading to substantial error reductions for both normal and abnormal traffic prediction: for delay prediction, AASTGCN reduces respectively MAE by 20.1\%, 21.2\%, and 6.6\%, and RMSE by 15.3\%, 17.0\%, and 9.9\% (relative to the top baseline methods). For queue prediction, MAE is reduced by 15.6\%, 18.6\%, and 12.2\%, and RMSE by 10.0\%, 12.6\%, and 13.7\%, respectively. Moreover, \cref{fig:11} reports performance across different prediction horizons from 15 min (short-term) to 60 min (longer-term), further confirming the superior performance of AASTGCN over the competing baselines. Notably, even under abnormal conditions, AASTGCN still maintains the lowest MAE and RMSE for both short-term predictions and longer-horizon delay and queue forecasting. 

\subsubsection{Effectiveness of historical features and abnormal experts}
The core innovations of AASTGCN lie in introducing historical features and the abnormal-aware expert mechanism. To validate their effectiveness, further ablation experiments are conducted using the same model structure but three variants: Only R, which uses only real-time features without abnormal expert (baseline);  R + H, which adds historical features but still trains a single model; R+H+A (original AASTGCN), which furthure includes the abnormal-aware expert to separately learn abnormal traffic patterns . \cref{tab:5} provides their prediction performances on the test set. From the results, we can conclude that:

\begin{itemize}[nosep]
    \item (\textit{R+H }\textit{vs.}\textit{ Only R}) Incorporating historical information reduces overall prediction errors and the gains mainly observed in normal cases (e.g., MAE drops by 10.5\% for delay) as normal traffic exhibits strong periodicity to use historical data key forecast references. In contrast, abnormal traffic often breaks periodic regularities, and thus R+H provides limited benefits or even underperform Only R for abnormal cases. 
    \item (\textit{R+H}\textit{+A}\textit{ vs. Only R}) Introducing the abnormal expert substantially improves performance for both normal and abnormal cases. For example, for queue prediction, MAE is further reduced by 18.9\% in normal cases and 12.5\% in abnormal conditions. It reveals that the abnormal-aware experts are necessary to disentangle abnormal and normal traffic patterns, thereby improving robustness to anomalies and enhancing overall accuracy.
\end{itemize}



\begin{table}[t]
    \centering
    \caption{Performance comparison of Only R, R+H, and R+H+A models.}
    \label{tab:5}
    \footnotesize\rmfamily
    \renewcommand{\arraystretch}{1.15}
    \setlength{\tabcolsep}{6pt}
    
    \begin{tabular*}{\textwidth}{@{\extracolsep{\fill}} llcccccc}
        \toprule
        \multirow{2}{*}{\textbf{Metric}} & \multirow{2}{*}{\textbf{Models\textsuperscript{*}}}
        & \multicolumn{2}{c}{\textbf{Overall}}
        & \multicolumn{2}{c}{\textbf{Normal}}
        & \multicolumn{2}{c}{\textbf{Abnormal}} \\
        \cmidrule(lr){3-4}\cmidrule(lr){5-6}\cmidrule(lr){7-8}
        & & MAE & RMSE & MAE & RMSE & MAE & RMSE \\
        \midrule
        
        \multirow[t]{3}{*}{\textbf{Delay (s)}} 
        & Only R
        & 4.376 & 8.492 & 4.175 & 7.548 & 11.398 & 24.518 \\
        & \makecell[l]{R + H\\\textit{(Improvement)}}
        & \makecell{3.964\\\textit{(-9.4\%)}}
        & \makecell{8.078\\\textit{(-4.9\%)}}
        & \makecell{3.738\\\textit{(-10.5\%)}}
        & \makecell{6.992\\\textit{(-7.4\%)}}
        & \makecell{11.832\\\textit{(--)}}
        & \makecell{25.248\\\textit{(--)}} \\
        & \makecell[l]{R + H + A\\\textit{(Improvement)}}
        & \makecell{\textbf{3.535}\\\textit{(-19.2\%)}}
        & \makecell{\textbf{7.332}\\\textit{(-13.7\%)}}
        & \makecell{\textbf{3.328}\\\textit{(-20.3\%)}}
        & \makecell{\textbf{6.409}\\\textit{(-15.1\%)}}
        & \makecell{\textbf{10.747}\\\textit{(-5.7\%)}}
        & \makecell{\textbf{22.295}\\\textit{(-9.1\%)}} \\
        \midrule
        
        \multirow[t]{3}{*}{\textbf{Queue (m)}}
        & Only R
        & 5.891 & 16.110 & 5.472 & 14.154 & 14.177 & 37.813 \\
        & \makecell[l]{R + H\\\textit{(Improvement)}}
        & \makecell{5.768\\\textit{(-2.1\%)}}
        & \makecell{15.839\\\textit{(-1.7\%)}}
        & \makecell{5.349\\\textit{(-2.2\%)}}
        & \makecell{13.574\\\textit{(-4.1\%)}}
        & \makecell{14.059\\\textit{(-0.8\%)}}
        & \makecell{39.598\\\textit{(--)}} \\
        & \makecell[l]{R + H + A\\\textit{(Improvement)}}
        & \makecell{\textbf{4.819}\\\textit{(-18.2\%)}}
        & \makecell{\textbf{13.878}\\\textit{(-13.9\%)}}
        & \makecell{\textbf{4.436}\\\textit{(-18.9\%)}}
        & \makecell{\textbf{12.186}\\\textit{(-13.9\%)}}
        & \makecell{\textbf{12.401}\\\textit{(-12.5\%)}}
        & \makecell{\textbf{32.630}\\\textit{(-13.7\%)}} \\
        \bottomrule
    \end{tabular*}
    
    \vspace{2pt}
    \begin{minipage}{0.98\linewidth}
        \footnotesize
        \emph{* R: Real-time features, H: historical features, A: abnormal-aware expert framework. Improvement is compared to the baseline: Only R.}
    \end{minipage}
\end{table}

\begin{figure}
    \centering
    \includegraphics[width=1\textwidth]{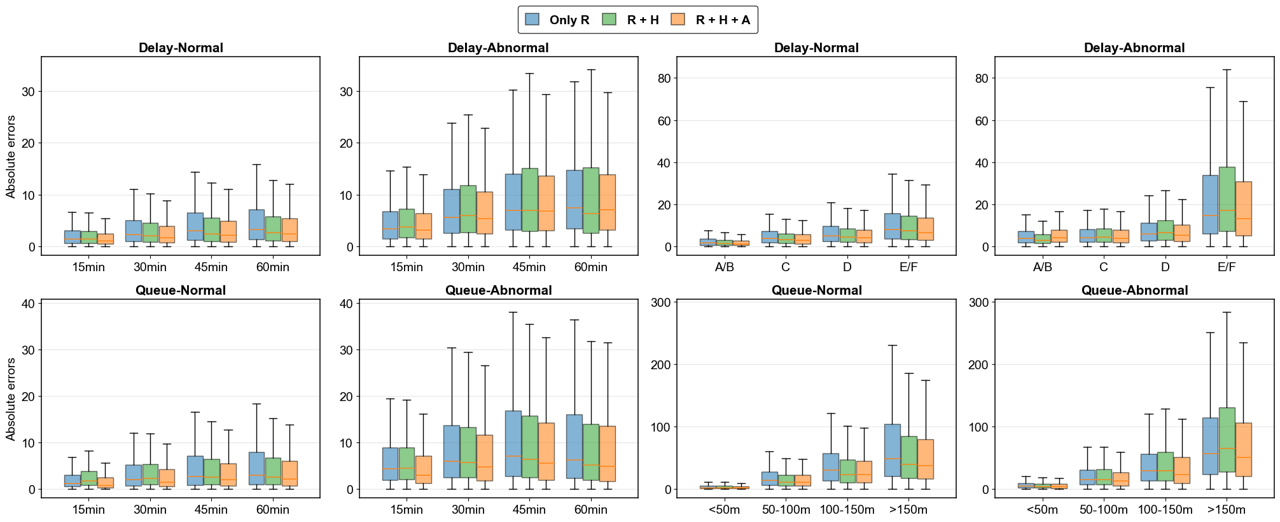}
    \vspace{-2em} 
    \caption{Model performance by prediction horizons and traffic conditions.}
    \label{fig:12}
\end{figure}

\cref{fig:12} visualizes their prediction errors by prediction horizons and traffic conditions. For delay, traffic LOS can be divided into levels A-F based on \citet{HCM2022}. Levels A/B (light traffic) and E/F (heavy congestion) are each merged due to limited samples. four levels of queue lengths are defined: <50m, 50-100m, 100-150m, and >150m. Results show a consistent trend that prediction errors increase with the forecasting horizon, and under abnormal cases are difficult to predict with higher errors than the normal cases. Across almost all horizons and traffic states, the full model (R+H+A) yields the lowest absolute prediction errors and the smallest error dispersion. Notably, the improvements are most pronounced under severe traffic conditions (i.e., LOS E/F for delay and levels C/D for queues). These findings confirm that the abnormal-aware expert mechanism is critical for robust forecasting under abrupt disruptions and heavy congestion, which are central challenges in real-world traffic management.  

\begin{figure}
    \centering
    \includegraphics[width=1\textwidth]{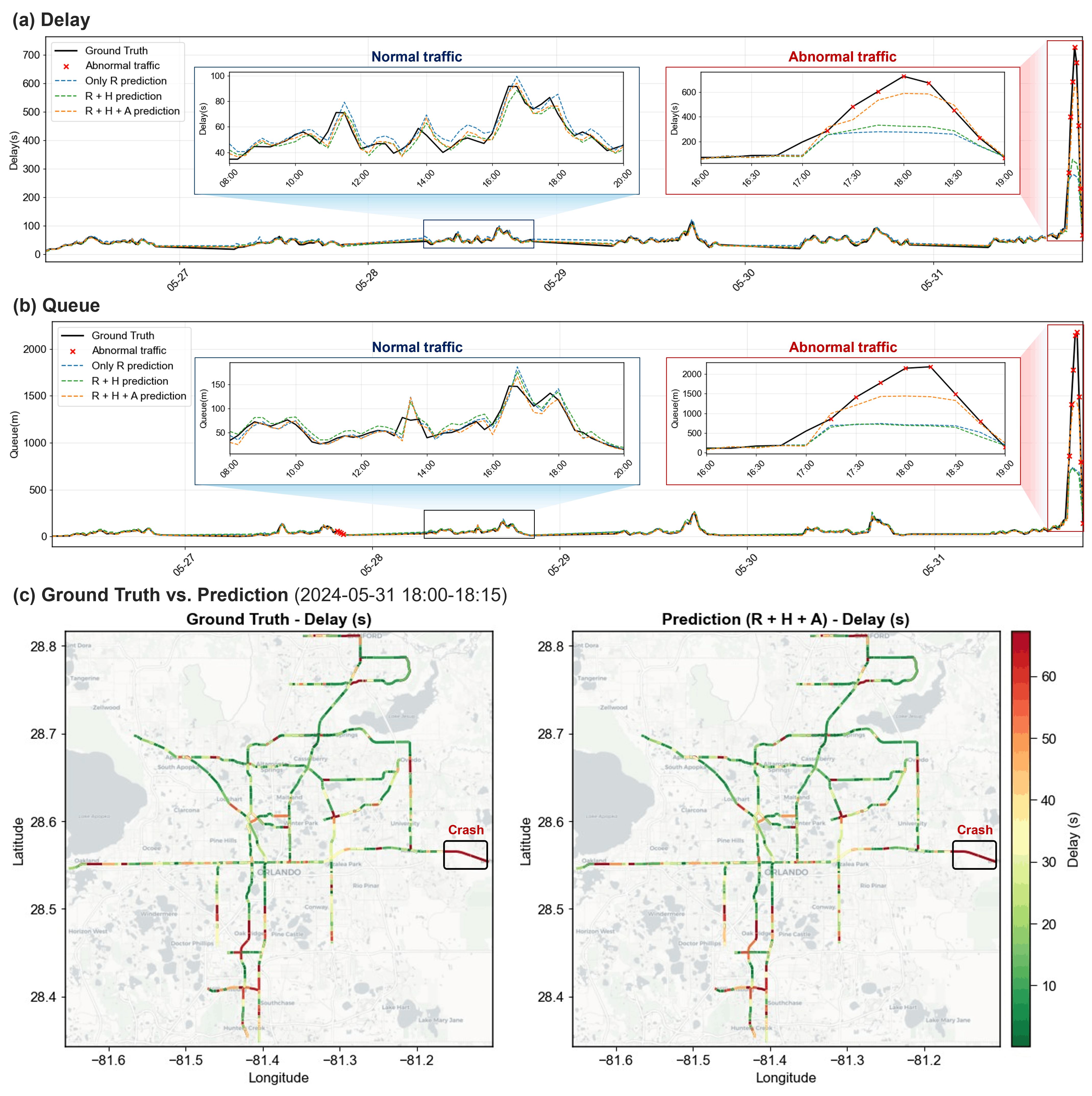}
    \vspace{-2em} 
    \caption{Model comparison case study.}
    \label{fig:13}
\end{figure}

\cref{fig:13} provide a case study to furthur compare the prediction abilities of three models. Using an arterial link on Colonial Dr in Orlando as an example, \cref{fig:13}(a)-(b) present the predicted and observed delay and queue from May-26 to May-31 in the test set. For the first five days, no abnormal events happen and both measures follow typical daily patterns. Within these normal traffic, three models show close predictions with the ground truth (black line), while predictions from Only R model (blue line) show more fluctuations. In contrast, model R+H and R+H+A yield stable predictions with smaller prediction errors. While in the afternoon (16:30-19:00) on May-31, a distinct abnormal episode occurs (a crash occured (\cref{fig:13}(c)) with significantly increased delay and queue. Under such abnormal traffic, only R+H+A model predicts the high delay and queue with similar time varing trend. However, the other two model fail to capture such abnormal conditions. \cref{fig:13}(c) compares network-level ground truth and R+H+A predictions, highlighting its strong spatiotemporal predictions at different traffic conditions (more in \cref{fig:A1}).

\begin{figure}
    \centering
    \includegraphics[width=0.9\textwidth]{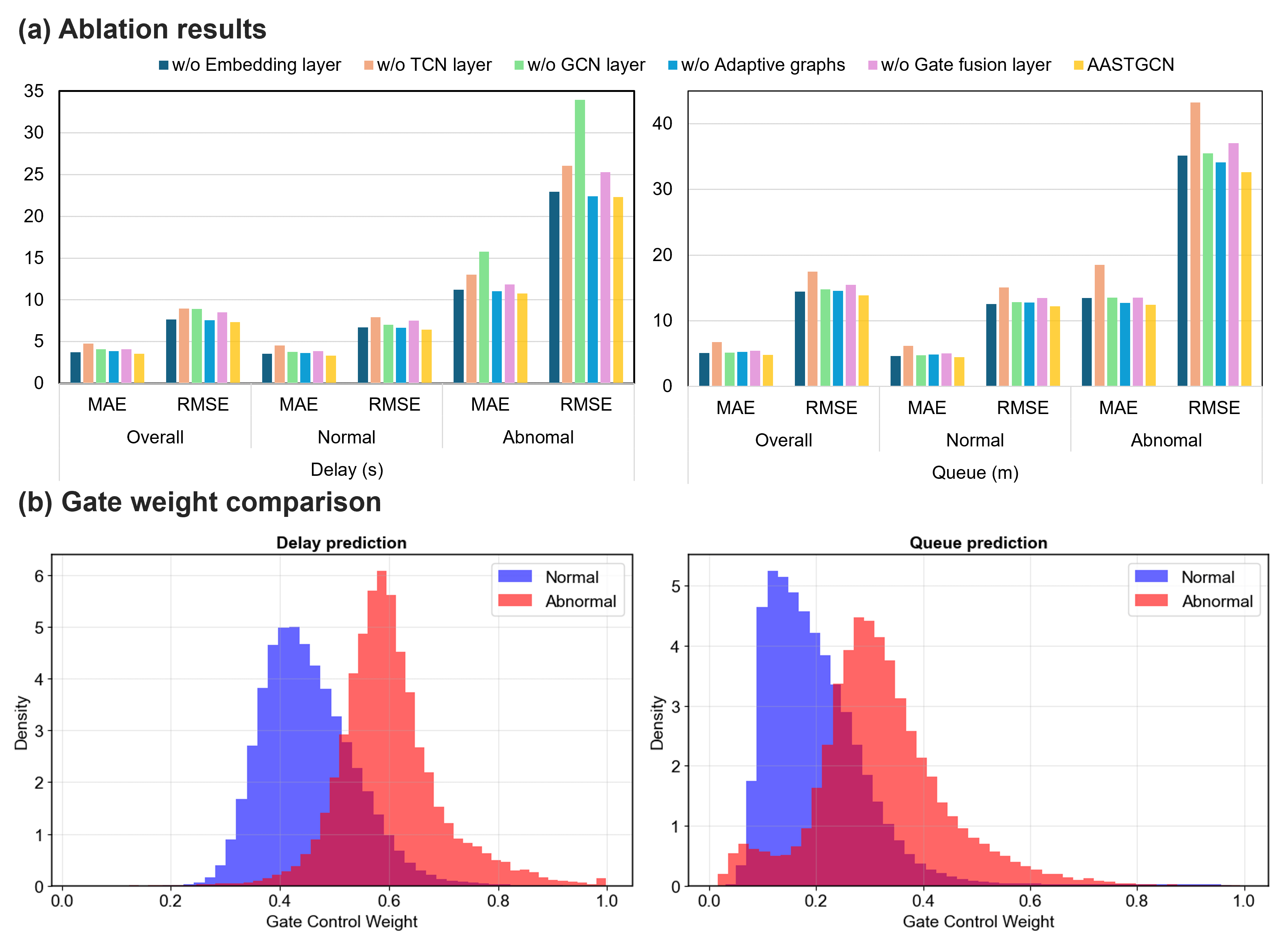}
    \vspace{-1em} 
    \caption{Model ablation results and gate weight comparison.}
    \label{fig:14}
\end{figure}

\subsubsection{Influence of AASTGCN components}
To assess the impact of key components within the AASTGCN framework, five variants are compared against the full AASTGCN (\cref{fig:14}(a)): (1) \textbf{w/o }\textbf{E}\textbf{mbedding}\textbf{ layer}, which\textbf{ }removes the input embedding module before ST-WaveNet; (2) \textbf{w/o }\textbf{TCN}\textbf{ layers},\textbf{ }which\textbf{ }replaces the TCN layer with simple MLPs to explicit temporal modeling; (3) \textbf{w/o }\textbf{GCN }\textbf{layers}, which\textbf{ }replaces the GCN layers with MLPs to exclude spatial modeling; (4) \textbf{w/o }\textbf{A}\textbf{daptive }\textbf{graphs}\textbf{, }which removes the adaptive adjacency matrix in the GCN module; (5) \textbf{w/o Gate}\textbf{ }\textbf{fusion}\textbf{ layer}, which\textbf{ }replaces the gated fusion mechanism to a concatenation layer. Overall, all variants produce higher errors than AASTGCN under both normal and abnormal conditions, confirming the effectiveness of these components. 

\textbf{1) }\textbf{Embedding layer and adaptive graphs.} Their impacts are moderate but consistent. Removing the embedding layer increases the overall MAE by 5.8\% (from 3.535 to 3.740) for delay and 5.0\% (from 4.819 to 5.060) for queue, suggesting that feature-specific embeddings help the model better represent heterogeneous inputs (e.g., dynamic traffic variables, temporal attributes, and static roadway characteristics). Removing the adaptive adjacency matrix increases the overall MAE by 8.8\% (3.535 to 3.846) for delay and 8.7\% (4.819 to 5.237) for queue, indicating that data-driven adaptive graphs provide additional spatial correlations beyond fixed topology, thereby improving prediction accuracy. 

\textbf{2) }\textbf{TCN and GCN}\textbf{ }\textbf{modules}\textbf{.}\textbf{ }These components have a substantially larger impact, highlighting their central role in spatiotemporal traffic modeling. Without TCN layers, the overall MAE increases by 34.7\% (3.535 to 4.763) for delay and 40.5\% (4.819 to 6.770) for queue, with consistently higher prediction errors across both normal and abnormal cases. It underscores that temporal modeling is critical, as future traffic is strongly dependent on recent traffic states. Similarly, removing the GCN layers leads to severe performance degradation, confirming that spatial dependency modeling is indispensable. Specifically, for delay prediction, MAE and RMSE increase by 46.5\% (10.747 to 15.749) and 52.4\% (22.295 to 33.971), respectively, demonstrating that the GCN module is crucial for capturing spatial correlations across arterial links, particularly for the delay prediction. 

\textbf{3) }\textbf{Gated fusion}\textbf{.}\textit{\textbf{ }}The gated fusion layer is also critical, as it adaptively balances real-time and historical information. Compared with the full AASTGCN, removing the gated fusion layer leads to consistently higher MAEs for both delay and queue prediction, with increases of 13.8\%–15.3\% under normal cases and 8.9\%–10.3\% under abnormal traffic. To further understand the gate fusion mechanism, \cref{fig:14}(b) visualizes the learned gate control weights , which controls the relative importance of real-time vs. historical features. Abnormal weights (red) are extracted from the abnormal export while normal weights (blue) are from the normal expert. A larger  means greater reliance on real-time traffic dynamics. The results show a clear regime-dependent pattern: under normal traffic cases, the gate weights are generally smaller, implying that the model leverages more historical-periodic information to improve predictions. Under abnormal conditions, the gate weights increase significantly, indicating that the model shifts higher attention toward real-time observations when traffic patterns deviate substantially from historical regularities.

\subsubsection{Sensitivity analysis of abnormal traffic detection}

\begin{figure}
    \centering
    \includegraphics[width=1\textwidth]{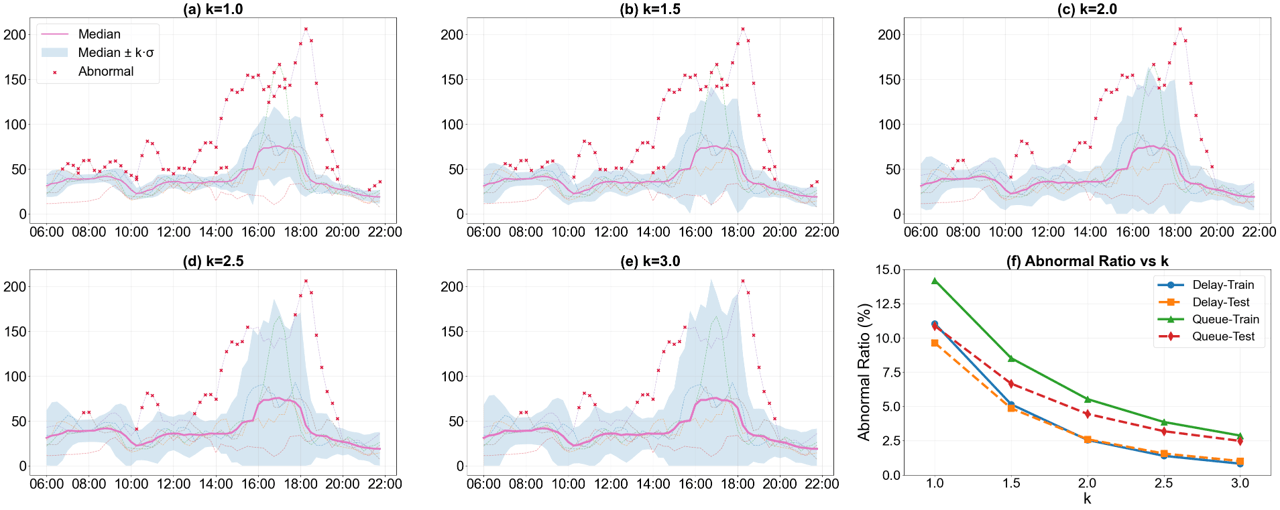}
    \vspace{-2em} 
    \caption{Abnormal traffic detection with different threshold $k$.}
    \label{fig:15}
\end{figure}

Since abnormal traffic is identified using the median-based method, the set of abnormal samples varies with the threshold $k$, which may further affect model prediction performance. As shown in \cref{fig:15}, a conservative (small) $k$ (i.e., 1 and 1.5) yields a narrower boundary and thus detects more observations as abnormal, whereas a larger $k$ (e.g., 2.5 and 3) produces more restrictive boundaries and detects fewer abnormal samples. Therefore, a sensitivity analysis is conducted to examine model performance under different values of $k$ ranging from 1 to 3 (i.e., 1, 1.5, 2, 2.5, and 3). \cref{fig:15}(f) shows the abnormal ratio in the modeling dataset, which decreases as $k$ increases in both the training and test sets. \cref{tab:6} summarizes the performance of models trained with different $k$ values and evaluated on the same datasets used previously. Overall, the results indicate stable performance across different $k$ values, with no significant differences for either queue or delay prediction. Models trained with $k \le 2$ achieve better performance for abnormal cases, likely because more abnormal samples are included during training. In summary, the dual-expert design of AASTGCN is robust to the choice of $k$, and a conservative threshold (i.e., $k \in [1,2]$) is recommended to capture more potential abnormal cases and further enhance performance under abnormal conditions.

\begin{table}
    \centering
    \caption{Sensitivity analysis performance.}
    \label{tab:6}
    \footnotesize\rmfamily
    \renewcommand{\arraystretch}{1.0}
    \setlength{\tabcolsep}{6pt}
    
    \begin{tabular*}{\textwidth}{@{\extracolsep{\fill}} llcccccc}
        \toprule
        \multirow{2}{*}{\textbf{Metric}} & \multirow{2}{*}{\textbf{Model}}
        & \multicolumn{2}{c}{\textbf{Overall}}
        & \multicolumn{2}{c}{\textbf{Normal}}
        & \multicolumn{2}{c}{\textbf{Abnormal}} \\
        \cmidrule(lr){3-4}\cmidrule(lr){5-6}\cmidrule(lr){7-8}
        & & MAE & RMSE & MAE & RMSE & MAE & RMSE \\
        \midrule
        
        \multirow[t]{5}{*}{\textbf{Delay (s)}} 
        & $k=1.0$ & 3.659 & 7.480 & 3.472 & 6.593 & 10.219 & 22.192 \\
        & $k=1.5$ & 3.589 & 7.379 & 3.389 & 6.402 & 10.555 & 22.918 \\
        & $k=2.0$ & \textbf{3.535} & \textbf{7.332} & \textbf{3.328} & 6.409 & 10.747 & \textbf{22.295} \\
        & $k=2.5$ & 3.662 & 7.457 & 3.449 & 6.522 & 11.132 & 22.637 \\
        & $k=3.0$ & 3.647 & 7.465 & 3.429 & 6.551 & 11.276 & 22.434 \\
        \midrule
        
        \multirow[t]{5}{*}{\textbf{Queue (m)}} 
        & $k=1.0$ & 4.954 & 14.073 & 4.560 & 12.462 & 12.733 & 32.304 \\
        & $k=1.5$ & 4.906 & \textbf{13.843} & 4.523 & 12.234 & 12.470 & \textbf{31.955} \\
        & $k=2.0$ & \textbf{4.819} & 13.878 & \textbf{4.436} & 12.186 & \textbf{12.401} & 32.630 \\
        & $k=2.5$ & 5.017 & 13.969 & 4.626 & \textbf{12.176} & 12.751 & 33.504 \\
        & $k=3.0$ & 5.461 & 14.205 & 5.058 & 12.418 & 13.439 & 33.800 \\
        \bottomrule
    \end{tabular*}
\end{table}

\section{Conclusion}
\label{sec:conclusion}
Emerging CV data shows great potential in arterial traffic state estimation and forecasting. Although existing studies have explored using CV data for arterial traffic (e.g., travel delay and queue) estimation and prediction, they mainly rely on the simulated high-penetration CV data and limited locations, which is hard to apply in real-world large arterial networks.  To address such gaps, we develop a novel CV data-based arterial network traffic state prediction framework with two components: (1) a two-stage traffic state extraction method which first estimates vehicle-level traffic measures and then aggregates them to obtain link-level traffic state measures. (2) an abnormality-aware spatiotemporal network AASTGCN which used a dual-expert architecture for both normal and abnormal traffic prediction and simultaneously models short-term traffic dynamics and long-term traffic periodic patterns. 

Real-world CV data are used to test the method in a large arterial network with 1,050 links. Experiment results show the effectiveness of the CV data-based traffic estimation method and the SOTA performance of the proposed AASTGCN. Based on the results, several conclusions can be summarized:
\begin{itemize}[nosep]
    \item The proposed traffic state extraction method is effective for large arterial networks (e.g., 1050 links in this study that are larger than prior studies) to provide real-time link-level traffic measures such as average travel delay, queue length, travel speed, which is very used for urban traffic operation monitoring and evaluation.  
    \item Based on historical traffic distribution, abnormal traffic can be detected in real-time, while it remains challenging for existing models to predict. By modeling abnormal cases separately with normal traffic in two dedicated expert networks, AASTGCN enhances prediction performance in both normal and abnormal cases. Meanwhile, its performance is robust to the choice of the abnormal traffic detection threshold.
    \item The gate-fusion mechanism in the AASTGCN model adaptively balances real-time and historical information for traffic prediction. It leverages historical-periodic information in normal traffic while shifting closer attention toward real-time traffic dynamics under abnormal traffic that deviates substantially from historical patterns.
\end{itemize}




Nonetheless, there are still a few limitations in the current study. First, the penetration rate of CV data is still relatively lower (2-3\%) at the current stage. The experiments are conducted based on such CV-based estimated measures, which may be biased by the driving behavior of connected vehicles. Higher penetration once available or multiple data fusion could be considered in future research. Second, this study mainly considers dynamic traffic and static road information; other factors, such as real-time weather and event information (e.g., work zones, road construction), which also affect traffic prediction, can be attempted in future efforts.

\section*{Acknowledgements}
This paper is funded partially by the Florida Department of Transportation under Grant BED26-977-18. FDOT assumes no liability for the contents or use thereof.

\section*{Authorship Contribution Statement}
\textbf{Lei Han}: Writing -- review \& editing; Writing -- original draft; Visualization; Validation; Methodology; Formal analysis; Data curation; Conceptualization.
\textbf{Mohamed Abdel-Aty}: Writing -- review \& editing; Methodology; Conceptualization; Supervision.
\textbf{Yang-Jun Joo}: Writing -- review \& editing; Conceptualization.

\section*{Declaration of Competing Interest}
The authors declare that they have no known competing financial interests or personal relationships.

\section*{Data Availability}
The authors do not have permission to share raw data. Processed data will be made available on request.

\appendix

\setcounter{table}{0}
\renewcommand{\thetable}{A.\arabic{table}}

\setcounter{figure}{0}
\renewcommand{\thefigure}{A.\arabic{figure}}

\section{Appendix}
\label{app:additional_results}

\begin{table}
    \centering
    \caption{Arterial link information.}
    \label{tab:additional_results}
    \footnotesize\rmfamily
    \renewcommand{\arraystretch}{1.2}
    \setlength{\tabcolsep}{5pt}
    \begin{tabularx}{\linewidth}{l c c c c X}
        \toprule
        \textbf{Arterial} 
        & \textbf{Length (mile)} 
        & \textbf{\# of links} 
        & \textbf{\# of intersections} 
        & \textbf{\# of lanes per direction} 
        & \textbf{Speed limits (mph)} \\
        \midrule
        25 ST              & 7.1  & 10  & 6  & [2, 3, 4]       & [35, 40, 50] \\
        ALAFAYA TRL        & 16.0 & 38  & 20 & [1, 2, 3, 4]    & [30, 35, 40, 45, 50] \\
        ALOMA AVE          & 8.6  & 26  & 14 & [2, 3, 4]       & [35, 40, 45] \\
        ALTAMONTE DR       & 5.2  & 24  & 13 & [3, 4]          & [40] \\
        CENTRAL FLA PKWY   & 7.8  & 20  & 11 & [2, 3, 4]       & [45] \\
        COLONIAL DR        & 62.9 & 166 & 84 & [2, 3, 4]       & [40, 45, 50, 55] \\
        EDGEWATER DR       & 2.7  & 4   & 3  & [2, 3, 4]       & [45] \\
        FAIRBANKS AVE      & 3.9  & 14  & 8  & [2, 3, 4]       & [35, 45] \\
        FOREST CITY RD     & 4.0  & 10  & 6  & [2, 3, 4]       & [45] \\
        FRENCH AVE         & 3.8  & 12  & 7  & [2, 3]          & [40, 45] \\
        JOHN YOUNG PKWY    & 34.4 & 88  & 45 & [3, 4]          & [35, 45, 55] \\
        KIRKMAN RD         & 12.1 & 36  & 19 & [2, 3, 4]       & [45, 50] \\
        LAKE MARY BLVD     & 21.3 & 52  & 27 & [2, 3, 4]       & [35, 45, 50] \\
        LEE RD             & 6.5  & 16  & 9  & [2, 3, 4]       & [35, 45] \\
        MILLS AVE          & 4.2  & 14  & 8  & [2, 3, 4]       & [35] \\
        ORANGE BLOSSOM TRL & 39.6 & 112 & 57 & [2, 3, 4]       & [35, 40, 45, 50, 55] \\
        ORLANDO AVE        & 7.6  & 30  & 16 & [2, 3, 4]       & [35, 40, 45] \\
        RED BUG LAKE RD    & 11.7 & 32  & 17 & [2, 3, 4]       & [45] \\
        SAND LAKE RD       & 5.1  & 14  & 8  & [3, 4]          & [45, 55] \\
        SEMORAN BLVD       & 26.8 & 80  & 41 & [3, 4]          & [45, 50] \\
        SR 436             & 11.9 & 42  & 22 & [3, 4]          & [35, 45] \\
        SR 46              & 8.5  & 20  & 11 & [2, 3, 4]       & [40, 45, 50, 55] \\
        SR-426             & 7.6  & 16  & 9  & [2, 3, 4]       & [30, 40, 45] \\
        SR-434-NS          & 5.5  & 16  & 9  & [2, 3, 4]       & [45] \\
        SR-434-WE          & 25.5 & 70  & 36 & [1, 2, 3, 4]    & [35, 40, 45, 50] \\
        US-17/92           & 20.6 & 52  & 27 & [2, 3, 4]       & [45, 50] \\
        US-441             & 15.5 & 36  & 19 & [1, 2, 3, 4]    & [35, 45, 55] \\
        
        \midrule
        \textbf{Total}     & 386.4 & 1050 & 521 & -- & -- \\
        \bottomrule
    \end{tabularx}
\end{table}

\begin{figure}
    \centering
    \includegraphics[height=0.95\textheight]{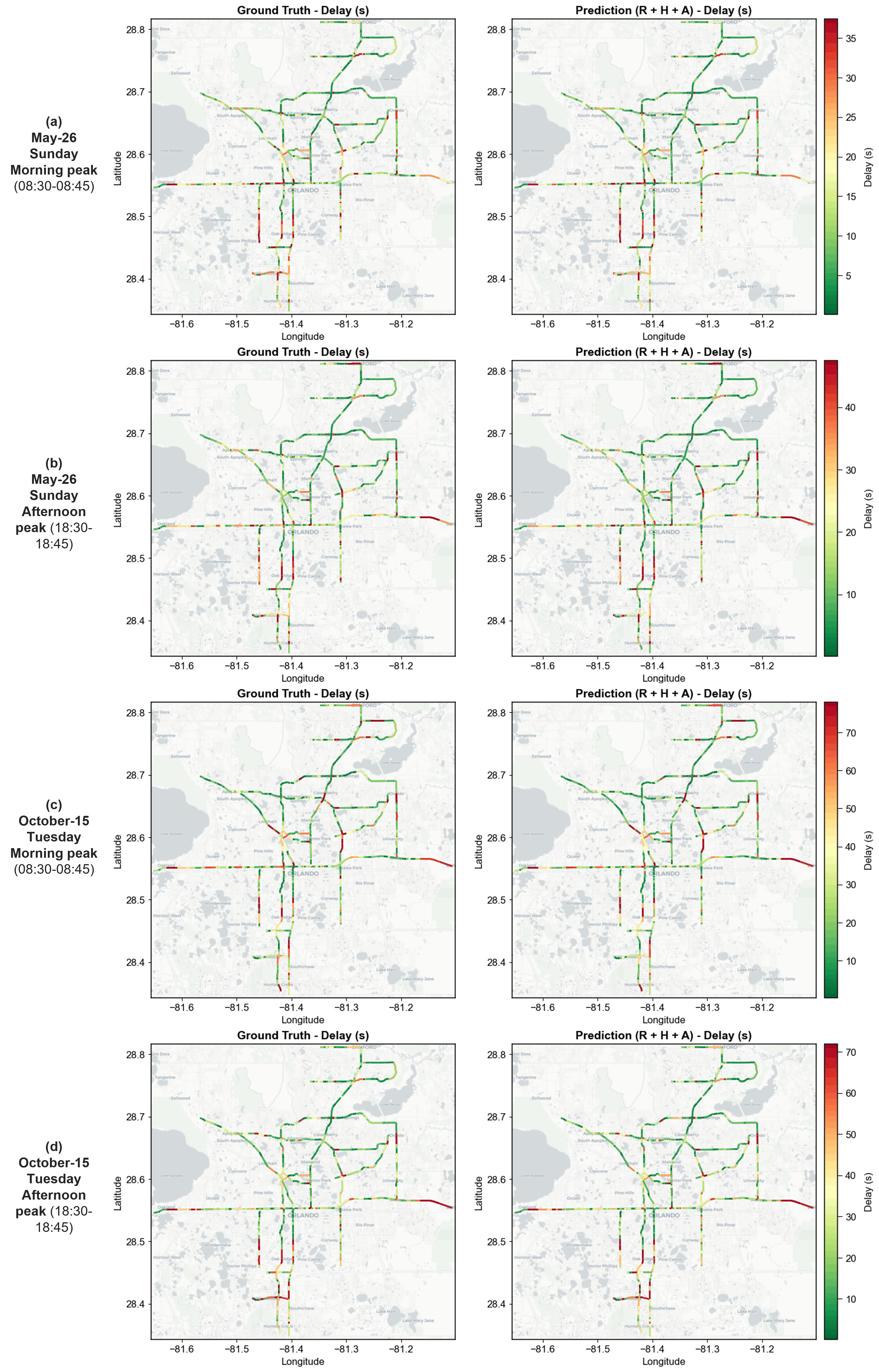}
    \vspace{0em} 
    \caption{Example of network traffic ground truth and predictions.}
    \label{fig:A1}
\end{figure}


\clearpage
\bibliography{refs}

\end{document}